\documentclass[11pt,letterpaper]{amsart}
\usepackage{amsmath,amssymb,amsthm,mathscinet}
\usepackage{enumerate}
\usepackage{tikz}
\usepackage{cite,hyperref}
\usepackage[noabbrev,capitalize]{cleveref}

\DeclareMathOperator{\RePart}{Re}
\DeclareMathOperator{\len}{len}
\DeclareMathOperator{\ord}{ord}
\DeclareMathOperator{\supp}{supp}
\DeclareMathOperator{\ADF}{ADF}
\DeclareMathOperator{\CDF}{CDF}
\DeclareMathOperator{\PSC}{PSC}
\DeclareMathOperator{\swap}{swap}
\DeclareMathOperator{\conjug}{conj}
\DeclareMathOperator{\crev}{crev}
\DeclareMathOperator{\srev}{srev}
\DeclareMathOperator{\scal}{scal}
\DeclareMathOperator{\subs}{subs}
\DeclareMathOperator{\Gol}{Gol}
\DeclareMathOperator{\EGol}{EGol}
\DeclareMathOperator{\SGol}{SGol}
\DeclareMathOperator{\UGol}{UGol}
\DeclareMathOperator{\BGol}{BGol}
\DeclareMathOperator{\RGol}{RGol}
\DeclareMathOperator{\RUGol}{RUGol}
\DeclareMathOperator{\RBGol}{RBGol}
\DeclareMathOperator{\GRS}{GRS}
\DeclareMathOperator{\SGI}{SGI}
\renewcommand{\Re}{\RePart}
\newcommand{\C}{{\mathbb C}}
\newcommand{\F}{{\mathbb F}}
\newcommand{\N}{{\mathbb N}}
\newcommand{\Z}{{\mathbb Z}}
\newcommand{\Ftwo}{\F_2}
\newcommand{\Zmz}{\Z\smallsetminus\{0\}}
\newcommand{\laur}{\C[z,z^{-1}]}
\newcommand{\conj}[1]{\overline{#1}}
\newcommand{\sums}[1]{\sum_{\substack{#1}}}
\newcommand{\norm}[2]{\Vert{#1}\Vert_{#2}}
\newcommand{\normp}[3]{\norm{#1}{#2}^{#3}}
\newcommand{\normt}[1]{\norm{#1}{2}}
\newcommand{\normtt}[1]{\normp{#1}{2}{2}}
\newcommand{\floor}[1]{\lfloor{#1}\rfloor}

\newtheorem{theorem}{Theorem}[section]
\newtheorem{proposition}[theorem]{Proposition}
\newtheorem{lemma}[theorem]{Lemma}
\newtheorem{corollary}[theorem]{Corollary}
\newtheorem{problem}[theorem]{Open Problem}
\theoremstyle{definition}
\newtheorem{definition}[theorem]{Definition}
\newtheorem{construction}[theorem]{Construction}
\newtheorem{example}[theorem]{Example}
\newtheorem{remark}[theorem]{Remark}

\title[Lowest Combined Autocorrelation and Crosscorrelation]{Sequence Pairs with Lowest Combined Autocorrelation and Crosscorrelation}
\author{Daniel J.~Katz}
\address{Department of Mathematics, California State University, Northridge, \: United States}
\author{Eli Moore}
\address{Department of Mathematics, California State University, Northridge, \: United States and Department of Mathematics, University of California, Davis, United States}
\thanks{This paper is based upon work of both authors supported in part by the National Science Foundation under Grant DMS-1500856 and upon work of Daniel J.~Katz supported in part by the National Science Foundation under Grant CCF-1815487.}
\date{04 March 2022}
\begin{document}
\begin{abstract}
Pursley and Sarwate established a lower bound on a combined measure of autocorrelation and crosscorrelation for a pair $(f,g)$ of binary sequences (i.e., sequences with terms in $\{-1,1\}$).
If $f$ is a nonzero sequence, then its autocorrelation demerit factor, $\text{ADF}(f)$, is the sum of the squared magnitudes of the aperiodic autocorrelation values over all nonzero shifts for the sequence obtained by normalizing $f$ to have unit Euclidean norm.
If $(f,g)$ is a pair of nonzero sequences, then their crosscorrelation demerit factor, $\text{CDF}(f,g)$, is the sum of the squared magnitudes of the aperiodic crosscorrelation values over all shifts for the sequences obtained by normalizing both $f$ and $g$ to have unit Euclidean norm.
Pursley and Sarwate showed that for binary sequences, the sum of $\text{CDF}(f,g)$ and the geometric mean of $\text{ADF}(f)$ and $\text{ADF}{(g)}$ must be at least $1$.
For randomly selected pairs of long binary sequences, this quantity is typically around $2$.
In this paper, we show that Pursley and Sarwate's bound is met for binary sequences precisely when $(f,g)$ is a Golay complementary pair.
We also prove a generalization of this result for sequences whose terms are arbitrary complex numbers.
We investigate constructions that produce infinite families of Golay complementary pairs, and compute the asymptotic values of autocorrelation and crosscorrelation demerit factors for such families.
\end{abstract}
\maketitle
\section{Introduction}
In this paper, we are interested in aperiodic correlation of sequences, so we define a {\it sequence} to be a doubly-infinite list of complex numbers, $f=(f_j)_{j \in \Z}=(\ldots,f_{-1},f_0,f_1,f_2,\ldots)$, in which only finitely many of the terms $f_j$ are nonzero.
The set of all sequences is a $\C$-vector space with the usual component-wise addition and $\C$-scalar multiplication.
The {\it support} of the sequence $f$, written $\supp(f)$, is the set of $j \in \Z$ such that $f_j\not=0$.
We say that a subset $S$ of $\Z$ is {\it contiguous} to mean that, whenever $S$ contains integers $a$ and $b$, it also contains every integer $c$ that lies between $a$ and $b$; the empty set is (vacuously) contiguous.
We say that a sequence $f$ is {\it contiguous} to mean that its support is contiguous.
If $f=(f_j)_{j \in \Z}$ is a sequence, then the {\it length} of $f$, written $\len f$, is the size of the smallest contiguous set that includes $\supp(f)$; so the length of the zero sequence is zero, and otherwise $\len f=\max\supp(f)-\min\supp(f)+1$.
Thus, $\len f=|\supp(f)|$ if $f$ is contiguous.
Certain types of sequences are especially interesting for applications.
A {\it unimodular sequence} $f$ is a contiguous sequence where $f_j$ is unimodular (i.e., $|f_j|=1$) for every $j \in \supp(f)$.
A unimodular sequence in which every nonzero term $f_j$ is an $m$th root of unity is called an {\it $m$-ary sequence}, and a {\it binary sequence} is a $2$-ary sequence, so that every nonzero term is in $\{-1,1\}$.
Although our main interest is in unimodular sequences, many of the constructions that are used to obtain unimodular sequences with good properties have non-unimodular sequences in intermediate steps, and this is why we must make these careful technical definitions and prove results that can handle sequences in general.

If $f=(f_j)_{j \in \Z}$ and $g=(g_j)_{j \in \Z}$ are sequences, then for $s \in \Z$, we define the {\it aperiodic crosscorrelation of $f$ with $g$ at shift $s$} to be
\begin{equation}\label{James}
C_{f,g}(s)=\sum_{j \in \Z} f_{j+s} \conj{g_j}.
\end{equation}
Only finitely many terms of the above sum can be nonzero due to the finite support of our sequences.
The {\it aperiodic autocorrelation of $f$ at shift $s$} is $C_{f,f}(s)$.
For the rest of this paper, we simply say {\it crosscorrelation} (resp., {\it autocorrelation}) to mean the aperiodic crosscorrelation (resp., aperiodic autocorrelation).

We write $\normt{f}$ for the Euclidean norm of $f$, i.e., $\normt{f}=\sqrt{\sum_{j \in \Z} |f_j|^2}$.
Then note that the autocorrelation at shift $0$ is is the squared Euclidean norm of $f$; that is,
\[
C_{f,f}(0)=\normtt{f},
\]
which is always a nonnegative real number, and is equal to $\len f$ if $f$ is unimodular.
If $f\not=0$ and  one wants a scaled version of $f$ that is a unit vector with respect to the Euclidean norm, then one should divide the terms of $f$ by $\normt{f}=\sqrt{C_{f,f}(0)}$ to obtain the {\it normalization of $f$}.
We say that a pair $(f,g)$ of sequences is {\it isoenergetic} to mean that $f$ and $g$ have the same Euclidean norm.
Many applications use isoenergetic pairs, which include pairs consisting of unimodular sequences of the same length.

In applications, one is interested in pairs $(f,g)$ of sequences where the crosscorrelation values of $f$ with $g$ at all shifts are small in magnitude, so that $f$ and $g$ are easily distinguished.
Furthermore, one wants the autocorrelation values of $f$ at all nonzero shifts to be small in magnitude, and similarly with $g$; this aids in synchronization.
There are two main ways that these goals of achieving smallness of correlation are measured: $l^\infty$ and $l^2$ methods.
The $l^\infty$ (worst-case) measures look at the largest magnitude among the undesirable correlations (i.e., autocorrelations of both sequences at nonzero shifts and all crosscorrelations between the two sequences).
The $l^\infty$ measure for autocorrelation of a sequence $f$ is called the {\it peak sidelobe level (PSL)}, and gives the maximum of $|C_{f,f}(s)|$ over all nonzero $s \in \Z$.
The $l^\infty$ measure for crosscorrelation for a pair $(f,g)$ is called the {\it peak crosscorrelation (PCC)}, and gives the maximum of $|C_{f,g}(s)|$ over all $s \in Z$.
The $l^2$ measures of smallness of correlation can be considered  mean square measures and are often called demerit factors (or, in reciprocal form, merit factors), and we shall define them in the next two paragraphs.
Pursley \cite{Pursley}, Burr \cite{Burr}, and K\"arkk\"ainen \cite{Karkkainen} all express the view that the $l^2$ measure is a better indicator of performance than the $l^\infty$ measure when evaluating the crosscorrelation of sequences in Code-Division Multiple Access (CDMA) applications.

For a pair $(f,g)$ of nonzero sequences, the {\it crosscorrelation demerit factor of $f$ with $g$} is defined by
\begin{equation}\label{Rupert}
\CDF(f,g) = \frac{\sum_{s \in \Z} |C_{f,g}(s)|^2}{C_{f,f}(0) C_{g,g}(0)}.
\end{equation}
This is the sum of squared magnitudes of all the crosscorrelation values for the normalization of $f$ with the normalization of $g$.
Only finitely many terms of the sum are nonzero since $C_{f,g}(s)=0$ whenever the shift $s$ is not a difference between an element of $\supp(f)$ and an element of $\supp(g)$.
Note that $\CDF(g,f)=\CDF(f,g)$ because $C_{g,f}(s)=\conj{C_{f,g}(-s)}$ for every $s \in \Z$.
The {\it crosscorrelation merit factor of $f$ with $g$} is the reciprocal of their crosscorrelation demerit factor.

For a nonzero sequence $f$, the {\it autocorrelation demerit factor of $f$} is defined by
\begin{equation}\label{Philip}
\ADF(f) = \frac{\sums{s \in \Z\smallsetminus\{0\}} |C_{f,f}(s)|^2}{C_{f,f}(0)^2} =-1+\CDF(f,f).
\end{equation}
This is the sum of squared magnitudes of all the autocorrelation values at nonzero shifts for the normalization of $f$.
The {\it autocorrelation merit factor of $f$} is the reciprocal of its autocorrelation demerit factor.

Since we want pairs $(f,g)$ of sequences with small magnitude autocorrelation values at nonzero shifts and small magnitude crosscorrelation values at all shifts, we want $\ADF(f)$, $\ADF(g)$, and $\CDF(f,g)$ all to be as small as possible.
Pursley and Sarwate \cite[eqs.~(3),(4)]{Pursley-Sarwate} proved bounds involving these three quantities when $f$ and $g$ are nonzero binary sequences of the same length; their bounds are
\begin{equation}\label{Mildred}
- \sqrt{\ADF(f)\ADF(g)} \leq \CDF(f,g)-1 \leq \sqrt{\ADF(f)\ADF(g)}.
\end{equation}
In particular, the lower bound shows that
\[
\sqrt{\ADF(f)\ADF(g)} + \CDF(f,g) \geq 1.
\]
This places a limitation on how low we can simultaneously make all three demerit factors.
Sarwate and Pursley later generalized their result to pairs of sequences with real terms \cite[eqs.~(8),(9)]{Sarwate-Pursley}.
In \cref{Gabriel} below, we show that Pursley and Sarwate's bounds hold for sequences whose terms are arbitrary complex numbers, and the sequences need not be of the same length.
This generalized bound could be obtained by going through Pursley and Sarwate's proof and making appropriate modifications, but we use a more efficient technique (based on Laurent polynomial representations of sequences) to obtain the key relations in full generality.

In view of this bound, we define the {\it Pursley--Sarwate criterion} of any pair $(f,g)$ of nonzero sequences to be
\[
\PSC(f,g)=\sqrt{\ADF(f)\ADF(g)} + \CDF(f,g),
\]
so that
\[
\PSC(f,g) \geq 1
\]
for all pairs $(f,g)$ of nonzero sequences.

Sarwate \cite[eqs.~(13),(38)]{Sarwate} showed that a random binary sequence $f$ of length $\ell$ (selected with uniform distribution) has an expected value for $\ADF(f)$ of $1-1/\ell$ and a randomly selected pair $(f,g)$ of such sequences has an expected value for $\CDF(f,g)$ of $1$.  So for pairs of randomly selected long binary sequences, we expect $\PSC(f,g)$ to be around $2$.

Since we want both autocorrelation and crosscorrelation to be as low as possible, we would like to know which pairs $(f,g)$ of sequences have $\PSC(f,g)=1$.
In fact, we find a complete classification of such pairs.
A pair of sequences $(f,g)$ with $C_{f,f}(s)+C_{g,g}(s)=0$ for all nonzero $s \in \Z$ is called a {\it Golay complementary pair} in honor of Golay who introduced them in \cite{Golay-51}.
It turns out that Golay complementarity is the key to achieving a Pursley--Sarwate criterion equal to $1$.
\begin{theorem}\label{Gabriel}
Let $f$ and $g$ be nonzero sequences.
Then
\[
-\sqrt{\ADF(f)\ADF(g)} \leq \CDF(f,g)-1 \leq \sqrt{\ADF(f)\ADF(g)}.
\]
Both these inequalities simultaneously become equalities if and only we have $\min \{\len f,\len g\}=1$, in which case $\PSC(f,g)=1$.
If $\min\{\len f,\len g\} > 1$, then equality is achieved in the lower bound (i.e., $\PSC(f,g)=1$) if and only if there is some $\lambda \in \C$ such that $(f,\lambda g)$ is a Golay complementary pair, in which case $\lambda\not=0$ and $(f,|\lambda| g)$ is also a Golay complementary pair.
In particular, if $\min\{\len f,\len g\}>1$ and $f$ and $g$ are unimodular, then equality is achieved in the lower bound if and only if $(f,g)$ is a Golay pair, in which case $\len f=\len g$ and $\ADF(f)=\ADF(g)$.
\end{theorem}
It should be noted that Pursley and Sarwate came close to this result, which comes from a Cauchy--Schwarz inequality.  In \cite{Pursley-Sarwate}, they state that the Cauchy--Schwarz inequality becomes an equality if and only if the two vectors are equal, and later in \cite{Sarwate-Pursley} they state that it becomes an equality if and only if the first vector is a scalar multiple of the second.  (The correct necessary and sufficient condition is that one of the vectors must be a scalar multiple of the other, so Pursley and Sarwate's second formulation is much closer to being correct than their first, but it still neglects the possibility that equality can occur when the second vector is zero and the first vector is nonzero.)  Pursley and Sarwate's first formulation of the necessary and sufficient condition for equality in the Cauchy--Schwarz inequality led them to state that it is impossible \cite[p.~305]{Pursley-Sarwate} to meet the lower bound in \eqref{Mildred}.  When they realized that this first formulation was incorrect and stated the second formulation, they noted \cite[p.~49]{Sarwate-Pursley} the unsoundness of their earlier argument that asserted the impossibility of meeting the lower bound in \eqref{Mildred}.
But their second paper still does not show that the lower bound is actually met, although they deduce the correct condition that would need to be met in the case of binary sequences.

Once \cref{Gabriel} is established, Turyn's construction \cite[Corollary to Lemma 5]{Turyn} supplies Golay pairs for infinitely many lengths, so we obtain infinitely many pairs $(f,g)$ of binary sequences with $\PSC(f,g)=1$.
\begin{corollary}
If $\ell=2^a 10^b 26^c$ for some nonnegative integers $a,b,c$, then there is a pair $(f,g)$ of binary sequences, with each sequence of length $\ell$, such that $\PSC(f,g)=1$.
\end{corollary}

\cref{Gabriel} tells us that the pairs $(f,g)$ of unimodular sequences of length greater than $1$ that have $\PSC(f,g)=1$ are precisely Golay complementary pairs.
We are interested in the relative magnitudes of $\ADF(f)$, $\ADF(g)$, and $\CDF(f,g)$ in Golay pairs.

Golay pairs are usually constructed by means of various transformations and combination rules starting from a small set of initial pairs.
We mention some of the simplest construction methods here.
If we have a sequence $f=(f_j)_{j \in  \Z}$ with $\supp(f)\subseteq \{0,1,\ldots,\len f-1\}$, then the {\it conjugate reverse of $f$}, written $f^\ddag$, is the sequence with $(f^\ddag)_j=\conj{f_{\len f-1-j}}$ for $j \in \{0,1,\ldots,\len f-1\}$ and $f^\ddag_j=0$ for all other $j$.
If both $f=(f_j)_{j \in \Z}$ and $g=(g_j)_{j\in\Z}$ are sequences of length $\ell$ whose supports are subsets of $\{0,1\ldots,\ell-1\}$, then the {\it concatenation of $f$ with $g$}, written $f|g$, is the sequence $c=(c_j)_{j \in \Z}$ of length $2\ell$ with $c_j=f_j$ for $j \in \{0,1,\ldots,\ell-1\}$, with $c_j=g_{j-\ell}$ for $j\in\{\ell,\ell+1,\ldots,2\ell-1\}$, and with $c_j=0$ for all other $j$.
The {\it interleaving of $f$ with $g$}, written $f\wr g$, is the sequence $h=(h_j)_{j \in \Z}$ of length $2\ell$ with $h_{2 j}=f_j$ and $h_{2 j+1}=g_j$ for $j \in \{0,1,\ldots,\ell-1\}$, and with $h_k=0$ for $k\not\in\{0,1\ldots,2\ell-1\}$.

The {\it Golay--Rudin--Shapiro recursion}, as typically employed, begins with a isoenergetic Golay pair ($f^{(0)},g^{(0)})$ (known as a {\it seed pair}), both of whose sequences are of the same positive length, $\ell$, and have supports that are subsets of $\{0,1,\ldots,\ell-1\}$.
Then the Golay--Rudin--Shapiro recursion produces a family $(f^{(n)},g^{(n)})_{n \geq 0}$ of Golay pairs by the rule $f^{(n+1)}=f^{(n)}|g^{(n)}$ and $g^{(n+1)}=f^{(n)}|-g^{(n)}$.
Thus, the length of sequences doubles at each step, so that $f^{(n)}$ and $g^{(n)}$ are sequences of length $2^n\ell$.
If the seed pair has sequences that are unimodular (resp., binary), then the entire family will consist of unimodular (resp., binary) sequences.
If one begins with the seed pair $(f^{(0)},g^{(0)})$ where both sequences are of length $1$ with their nonzero terms equal to $1$, then the construction produces the Rudin--Shapiro sequences.
One consequence of our \cref{Sally} is that we know the asymptotic behavior of the demerit factors for such a family.
\begin{theorem}\label{Jean}
Let $(f^{(n)},g^{(n)})_{n \geq 0}$ be a family of Golay pairs produced by the Golay--Rudin--Shapiro recursion as described above.  Then
 \begin{align*} \lim_{n\to\infty} \ADF(f^{(n)})=\lim_{n\to\infty} \ADF(g^{(n)}) & =1/3, \text{ and}\\
\lim_{n\to\infty} \CDF(f^{(n)},g^{(n)}) & =2/3.
\end{align*}
\end{theorem}

The {\it simple Golay interleaving recursion}, as typically employed, also begins with a seed Golay pair ($f^{(0)},g^{(0)})$ that is isoenergetic, with both sequences of the same positive length, $\ell$, and having supports that are subsets of $\{0,1,\ldots,\ell-1\}$.
The simple Golay interleaving recursion produces a family $(f^{(n)},g^{(n)})_{n \geq 0}$ of Golay pairs by the rule $f^{(n+1)}=f^{(n)}\wr g^{(n)}$ and $g^{(n+1)}=g^{(n)\ddag}\wr-f^{(n)\ddag}$.
Thus the length of sequences doubles at each step, so that $f^{(n)}$ and $g^{(n)}$ are sequences of length $2^n\ell$.
If the seed pair has sequences that are unimodular (resp., binary), then the entire family will consist of unimodular (resp., binary) sequences.
Our \cref{Griffin} shows that such families have the same asymptotic behavior as those produced by the Golay--Rudin--Shapiro recursion.
\begin{theorem}\label{Emily}
Let $(f^{(n)},g^{(n)})_{n \geq 0}$ be a family of Golay pairs produced by the simple Golay interleaving recursion as described above.  Then
\begin{align*} \lim_{n\to\infty} \ADF(f^{(n)})=\lim_{n\to\infty} \ADF(g^{(n)}) & =1/3, \text{ and}\\
\lim_{n\to\infty} \CDF(f^{(n)},g^{(n)}) & =2/3.
\end{align*}
\end{theorem}

In fact, our Theorems \ref{Sally} and \ref{Griffin} show that we obtain the same asymp\-tot\-ic behavior described in Theorems \ref{Jean} and \ref{Emily} here even if we allow ourselves much greater freedom in constructing the infinite families of pairs than indicated here.
This freedom comes by applying a transformation to each pair $(f^{(n)},g^{(n)})$ before applying the recursion rule that doubles the length of the sequences.
These transformations are inspired by the work of Golay, who showed that there are $64$ transformations of a binary pair $(f,g)$ that preserve the Golay complementary property: these include exchanging $f$ for $g$, negating $f$, negating $g$, reversing $f$, reversing $g$, negating the terms with odd indices in both $f$ and $g$, and compositions of any selection of these transformations.
These transformations also work for Golay pairs with complex-valued sequences, provided that one uses the conjugate reverse operation as the generalization of Golay's reverse operation.
\cref{Sally} shows that one still gets the same limiting values for demerit factors as attested in \cref{Jean} if one applies such transformations at each stage of the Golay--Rudin--Shapiro recursion before the next doubling of length by concatenation (and one may use different transformations at different stages).
\cref{Griffin} shows that the interleaving construction also exhibits the same limiting behavior as in \cref{Emily} when one uses the transformations between doublings, although it places the restriction that whenever one conjugate reverses the first sequence in a pair, one must also conjugate reverse the second sequence.
Even now, we have not indicated the full scope of Theorems \ref{Sally} and \ref{Griffin}, which allow for more general choices of seed Golay pairs and transformations to be used at each step.
It should also be noted that Theorems \ref{Sally} and \ref{Griffin} provide exact formulae for $\ADF(f^{(n)})$, $\ADF(g^{(n)})$, and $\CDF(f^{(n)},g^{(n)})$ for each $n$, from which one obtains the asymptotic results given in Theorems \ref{Jean} and \ref{Emily} here.

This paper is organized as follows.
\cref{Nancy} introduces the formalism of viewing sequences as Laurent polynomials, and provides the notations and conventions for the rest of the paper, along with proofs of some preliminary facts.
\cref{Henry} is the proof of \cref{Gabriel}.
\cref{Leonard} describes a group that generalizes Golay's collection of $64$ transformations that preserve complementarity of pairs; this group is very useful in our proofs on the asymptotic behavior of demerit factors.
\cref{Clarence} describes a single construction for complex Golay pairs that is more general and simpler to use than previous constructions of Golay and Turyn; this eases our proofs of asymptotic results.
\cref{Taylor} has the proof \cref{Sally} (of which \cref{Jean} here is one corollary) on the behavior of demerit factors for families produced by the Golay--Rudin--Shapiro recursion.
\cref{Valerie} has the proof \cref{Griffin} (of which \cref{Emily} here is one corollary) on the behavior of demerit factors for families produced by the simple Golay interleaving recursion.
\cref{Zeke} poses an open problem that asks whether all families of Golay pairs consisting of binary sequences whose lengths tend to infinity have asymptotic autocorrelation and crosscorrelation demerit factors that tend to $1/3$ and $2/3$, respectively.
The same question is also asked of the more general class of Golay pairs consisting of unimodular sequences.

\section{Preliminaries}\label{Nancy}

Recall from the Introduction the definition of a sequence, its support, its length, its Euclidean norm, and its normalization.
Also recall what it means for a subset of $\Z$ or a sequence to be contiguous, as well as the definitions of contiguous, unimodular, $m$-ary, and binary sequences, and of isoenergetic sequence pairs.
We also continue to use the definitions of crosscorrelation and autocorrelation, with their respective demerit factors, and the Pursley--Sarwate criterion.
One should also recall the definition of Golay complementary pairs, which are also simply called {\it Golay pairs} or {\it complementary pairs}.
All these definitions and their notations from the Introduction remain in force throughout this paper.
We use $\N$ to denote the set $\{0,1,\ldots\}$ of nonnegative integers.

We identify the sequence $f=(f_j)_{j \in \Z}=(\ldots,f_{-1},f_0,f_1,f_2,\ldots)$ with the Laurent polynomial $f(z)=\sum_{j \in \Z} f_j z^j$ in $\C[z,z^{-1}]$, so that the definitions and notations for support, contiguity, length, unimodularity, $m$-arity, binarity, Euclidean norm, being isoenergetic, correlation, demerit factors, the Pursley--Sarwate criterion, and Golay complementarity all apply equally well to Laurent polynomials.
We also use the convention that any property defined for a sequence (Laurent polynomial) can be predicated of a pair, in which case this is understood to mean that both sequences in the pair have that property.
So, for example, a unimodular Golay pair $(f,g)$ with $\len(f,g)=10$ is a Golay pair $(f,g)$ where $f$ and $g$ are both unimodular and $\len f=\len g=10$.
A {\it monomial} is sequence of length $1$, that is, some $c z^j$ where $c$ is a nonzero complex number and $j \in \Z$.

If $f(z) \in \laur$ is a nonzero Laurent polynomial, then its {\it order}, written $\ord f$, is the smallest $j$ such that $f_j\not=0$, while its {\it degree}, written $\deg f$, is the largest $k$ such that $f_k\not=0$.
We create two new symbols, $\infty$ and $-\infty$, and decree that $\ord 0=\infty$ and $\deg 0=-\infty$.
We extend the addition operation from $\Z$ to $\Z\cup\{\infty,-\infty\}$ by the following rules: (I) $a+\infty=\infty+a=\infty$ for $a \in \Z\cup\{\infty\}$, (II) $b+(-\infty)=(-\infty)+b=-\infty$ for $b \in \Z\cup\{-\infty\}$, and (III) $(-\infty)+\infty=\infty+(-\infty)=0$.
For every $f,g \in \laur$, the first rules makes $\ord(f g)=\ord f+\ord g$, the second rule makes $\deg(f g)=\deg f+\deg g$, and the third rule makes $\ord f(z)+\deg f(z^{-1})=0$.
Note that $\len f=\deg f-\ord f+1$ for every nonzero $f$ in $\laur$.

If $f(z)=\sum_{j \in \Z} f_j z^j\in \laur$, then we write $\conj{f(z)}$ to mean $\sum_{j \in \Z} \conj{f_j} z^{-j}$, where the inversion of $z$ comes about because we are interested in our polynomials on the complex unit circle.
We sometimes use $f$ as a shorthand for $f(z)$, and in this case $\conj{f}$ is a shorthand for $\conj{f(z)}$.
We do not need or introduce any notation for the operation that simply conjugates the coefficients of a Laurent polynomial without also inverting the indeterminate.
We also use the convention that if $f(z) \in \laur$, then $|f(z)|^2=f(z) \conj{f(z)}$, and indeed, if $k$ is any nonnegative integer, then $|f(z)|^{2 k}=(f(z) \conj{f(z)})^k$.
When we abbreviate $f(z)$ by $f$, then $|f|^2$ and $|f|^{2 k}$ stand for $|f(z)|^2$ and $|f(z)|^{2 k}$, respectively.
Along the same lines, if $f(z) \in \laur$, then $\Re f(z)$ is a shorthand for $(f(z)+\conj{f(z)})/2$ (which can be abbreviated $\Re f=(f+\conj{f})/2$).

For any $f(z)\in\laur$, we use the convention that $f_s$ is the coefficient of $z^s$ in $f(z)$.
Sometimes we use enclosing parentheses when the Laurent polynomial has a complicated form, for example if $f(z)$ and $g(z)$ are Laurent polynomials, then $(fg)_s$ is the coefficient of $z^s$ in the product $f(z)g(z)$.
If we write $f^k_s$, then we mean the coefficient of $z^s$ in $f(z)^k$, that is, we mean $(f^k)_s$.
And similarly $|f|^{2 k}_s$ means the coefficient of $z^s$ in $|f(z)|^{2 k}$, that is $(|f|^{2 k})_s$.
Thus if $c \in \C$ and $n$ is a nonzero integer, then $f(c z^n)_{n m}=c^m f_m$ for every $m \in \Z$; in particular $f(c z^n)_0 =f_0$.
If, in addition, $c$ is unimodular, then $|f(c z^n)|^2_{m n} = c^m |f|^2_m$ for every $m \in \Z$; in particular $|f(c z^n)|^2_0=|f|^2_0$.

We now show how the correlation concepts, defined for sequences in the Introduction, are realized in the Laurent polynomial interpretation.
If $f(z), g(z) \in \laur$, we define the {\it crosscorrelation of $f$ with $g$ at shift $s$} to be
\begin{equation}\label{Eric}
C_{f,g}(s)=(f\conj{g})_s,
\end{equation}
which agrees with \eqref{James} when $f$ and $g$ are Laurent polynomials representing sequences.
So the Laurent polynomial $f\conj{g}$ records all the crosscorrelation values:
\begin{equation}\label{Edwin}
f\conj{g} = \sum_{s \in \Z} C_{f,g}(s) z^s.
\end{equation}
When $f=g$, we call $C_{f,f}(s)$ the {\it autocorrelation of $f$ at shift $s$}.  Note that
\begin{equation}\label{Dorothy}
C_{f,f}(0)=|f|^2_0 = \sum_{s \in \Z} |f_s|^2=\normtt{f},
\end{equation}
which is the squared Euclidean norm of $f$.
Thus, a pair $(f,g) \in \laur^2$ is isoenergetic if and only if $C_{f,f}(0)=C_{g,g}(0)$.
We record some basic facts about correlation.
\begin{lemma}\label{Samuel}
Let $f(z), g(z) \in \laur$ and $s \in \Z$.
\begin{enumerate}[(i).]
\item\label{Theodore} $C_{f,f}(0)=|f|^2_0=\normtt{f}=0$ if and only if $f=0$; otherwise $C_{f,f}(0)$ is positive real.
\item\label{Quentin} If $f$ is unimodular, then $C_{f,f}(0)=|f|^2_0=\normtt{f}=\len f$.
\item\label{Ulrich} $C_{g,f}(s)=\conj{C_{f,g}(-s)}$.
\item\label{Vanessa} $C_{f,f}(s)=\conj{C_{f,f}(-s)}$.
\item\label{Wilfrid} If $f,g\not=0$ and either $s > \deg f-\ord g$ or $s < \ord f-\deg g$, then $C_{f,g}(s)=0$.
\item\label{Xavier} If $f,g\not=0$, then $C_{f,g}(\deg f-\ord g)=f_{\deg f} \conj{g_{\ord g}}\not=0$ and $C_{f,g}(\ord f-\deg g)=f_{\ord f} \conj{g_{\deg g}}\not=0$.
\item\label{Yolanda} If $|s| \geq \len f$, then $C_{f,f}(s)=0$.
\item\label{Zosimos} If $f\not=0$, then $C_{f,f}(\len f-1)=f_{\deg f} \conj{f_{\ord f}}\not=0$ and $C_{f,f}(1-\len f)=f_{\ord f} \conj{f_{\deg f}}\not=0$.
\end{enumerate}
\end{lemma}
\begin{proof}
One immediately obtains part \eqref{Theodore} from \eqref{Dorothy}, and part \eqref{Quentin} also follows from \eqref{Dorothy} since the sum of the squared magnitudes of $\len f$ unimodular numbers is $\len f$.
From \eqref{Eric}, we see that $\conj{C_{f,g}(-s)}=\conj{(f\conj g)_{-s}}=(\conj{f\conj g})_s=(g\conj f)_s$, which by another application of \eqref{Eric} is $C_{g,f}(s)$.
This proves part \eqref{Ulrich}, and if one sets $g=f$ there, then part \eqref{Vanessa} follows.
If $f,g\not=0$, then the highest degree terms in $f$ and $\conj{g}$ are $f_{\deg f} z^{\deg f}$ and $\conj{g_{\ord g}} z^{-\ord g}$, respectively, while the lowest degree terms in $f$ and $\conj{g}$ are $f_{\ord f} z^{\ord f}$ and $\conj{g_{\deg g}} z^{-\deg g}$, respectively.
Thus \eqref{Eric} shows that $C_{f,g}(s)=0$ if $s > \deg f-\ord g$ or $s < \ord f-\deg g$, and shows that $C_{f,g}(\deg f-\ord g)$ and $C_{f,g}(\ord f-\deg g)$ have the desired values.
This proves part \eqref{Wilfrid} and part \eqref{Xavier}, and if one sets $g=f$ in these and recalls that $\len f = \deg f-\ord f+1$ and notes that $C_{0,0}(s)=0$ for all $s \in \Z$, then one obtains part \eqref{Yolanda} and part \eqref{Zosimos}.
\end{proof}

We can use \eqref{Edwin} and \eqref{Dorothy} together to obtain
\begin{equation}\label{Matilda}
|fg|^2_0 = \sum_{s \in \Z} |C_{f,g}(s)|^2.
\end{equation}
Then comparing expressions \eqref{Matilda} and \eqref{Dorothy} with the terms in the definition \eqref{Rupert} of the crosscorrelation demerit factor, we see that if $f,g\not=0$, then
\begin{equation}\label{Bart}
\CDF(f,g)=\frac{|f g|^2_0}{|f|^2_0 |g|^2_0},
\end{equation}
and thus by \eqref{Philip}
\begin{equation}\label{Vera}
\ADF(f)=-1+\frac{|f|^4_0}{(|f|^2_0)^2}.
\end{equation}
Expressions \eqref{Bart} and \eqref{Vera} connect crosscorrelation and autocorrelation demerit factors to $L^p$ norms of polynomials on the complex unit circle; see \cite[p.~515--516]{Katz-Lee-Trunov-a} for more details.
We record a simple criterion for vanishing autocorrelation demerit factor.
\begin{lemma}\label{Eugene}
Let $f$ be a nonzero sequence.
Then $\ADF(f)=0$ if and only if $\len f=1$.
\end{lemma}
\begin{proof}
Parts \eqref{Yolanda} and \eqref{Zosimos} of \cref{Samuel} show that $C_{f,f}(s)=0$ for all nonzero $s$ if and only if $\len f\leq 1$.
Since $f$ is nonzero, this means that the numerator of $\ADF(f)$ in the expression in \eqref{Philip} is zero if and only if $\len f=1$.
\end{proof}
The following lemma translates the definition of Golay complementary pair to the Laurent polynomial interpretation.
\begin{lemma}\label{Allison}
If $(f,g) \in \laur^2$, then $(f,g)$ is a Golay pair if and only if $|f|^2 + |g|^2$ is constant, in which case $|f|^2 + |g|^2 =|f|^2_0+|g|^2_0$.
\end{lemma}
\begin{proof}
By \eqref{Edwin}, we have $|f|^2 + |g|^2 = \sum_{s \in \Z} (C_{f,f}(s) + C_{g,g}(s)) z^{s}$, which is constant if and only if $C_{f,f}(s) + C_{g,g}(s) = 0$ for all nonzero $s$, that is, if and only if $(f,g)$ is a Golay pair.
And $|f|^2+|g|^2$ is constant if and only if it equals its own constant term, which is $|f|^2_0+|g|^2_0$.
\end{proof}
We should note that, with trivial exceptions, sequences that form a Golay pair must be the same length.
\begin{lemma}\label{Felix}
If $(f,g)$ is a Golay pair, then either $\len f=\len g$ or else $\{\len f,\len g\}=\{0,1\}$.
\end{lemma}
\begin{proof}
Suppose that $(f,g)$ is a Golay pair.
If one sequence has a length $\ell$ with $\ell > 1$, then \cref{Samuel}\eqref{Zosimos} shows that its autocorrelation at shift $\ell-1$ is nonzero, so that the Golay condition forces the other sequence to have nonzero autocorrelation at shift $\ell-1$, which by \cref{Samuel}\eqref{Yolanda} forces it to have length at least $\ell$.
Thus, if either sequence has length greater than $1$, they both must have the same length.
\end{proof}
We should also note that in many Golay pairs of interest, the autocorrelation demerit factors of the two elements are the same.
\begin{lemma}\label{Horatio}
Let $(f,g)$ be a Golay pair with $f,g\not=0$.
If $(f,g)$ is isoenergetic, then $\ADF(f)=\ADF(g)$.
If $\ADF(f)=\ADF(g)$, then $(f,g)$ is isoenergetic or $\len f=\len g=1$.
\end{lemma}
\begin{proof}
Since $(f,g)$ is a Golay pair, we have $C_{f,f}(s)=-C_{g,g}(s)$ for every nonzero $s$, and so $\sum_{s \in \Z\smallsetminus\{0\}} |C_{f,f}(s)|^2=\sum_{s \in \Z\smallsetminus\{0\}} |C_{g,g}(s)|^2$.
Thus, if $(f,g)$ is isoenergetic, then $C_{f,f}(0)=C_{g,g}(0)$, and then \eqref{Philip} shows that $\ADF(f)=\ADF(g)$.
Conversely, if $\ADF(f)=\ADF(g)$, then \eqref{Philip} shows that either $C_{f,f}(0)=C_{g,g}(0)$ (so that $(f,g)$ is isoenergetic) or $\ADF(f)=\ADF(g)=0$, the latter of which can only hold if $\len f=\len g=1$ by \cref{Eugene}.
\end{proof}
Unimodular Golay pairs are of particular interest and have many of the good properties discussed above.
\begin{lemma}\label{Felicia}
If $f,g$ are nonzero unimodular sequences such that $(f,g)$ is a Golay pair, then $\len f=\len g=\normtt{f}=\normtt{g}$ (so that $(f,g)$ is isoenergetic), $|f|^2+|g|^2=2\len f=2\len g$, and $\ADF(f)=\ADF(g)$.
\end{lemma}
\begin{proof}
\cref{Felix} shows that $\len f=\len g$.
Then \cref{Samuel}\eqref{Quentin} shows that $\normtt{f}=|f|^2_0=\len f = \len g= |g|_0^2=\normtt{g}$, so that $(f,g)$ is isoenergetic, and then \cref{Allison} shows that $|f|^2+|g|^2=2\len f$, while \cref{Horatio} shows that $\ADF(f)=\ADF(g)$.
\end{proof}

\section{Proof of \cref{Gabriel}}\label{Henry}

We now prove our first main result, \cref{Gabriel}.
If $\len f=1$ (resp., $\len g=1$), then \cref{Eugene} shows that $\ADF(f)=0$ (resp., $\ADF(g)=0$) and one easily calculates $\CDF(f,g)=1$ from \eqref{Bart}, so that both inequalities are achieved simultaneously and we have $\PSC(f,g)=1$.
So henceforth we assume that $\min\{\len f,\len g\}>1$.

Let $\Gamma=(f g \conj{f g})_0$, which we can interpret in two different ways.
If we group the terms as $\Gamma=(|f|^2 \conj{|g|^2})_0$, and use \eqref{Edwin} to interpret both $|f|^2=f\conj{f}$ and $|g|^2=g\conj{g}$ as sequences whose terms are autocorrelation values, then \eqref{Eric} shows that $\Gamma$ is the crosscorrelation of the autocorrelation spectrum of $f$ with the autocorrelation spectrum of $g$ at shift $0$, i.e., $\Gamma=\sum_{s \in \Z} C_{f,f}(s) \conj{C_{g,g}(s)}$.
If instead we choose to group the terms as $\Gamma=|f\conj{g}|^2_0$, then \eqref{Dorothy} shows that $\Gamma$ is the sum of the squared magnitudes of the coefficients of $f\conj{g}$, so by \eqref{Edwin} we have $\Gamma=\sum_{s \in \Z} |C_{f,g}(s)|^2$, which shows that $\Gamma$ is real.
Now define $\Delta=-C_{f,f}(0) C_{g,g}(0)+\Gamma$.
Since $\Gamma$, $C_{f,f}(0)$, and $C_{g,g}(0)$ are real, $\Delta$ is a real number with
\begin{equation}\label{Irene}
\Delta=-C_{f,f}(0) C_{g,g}(0) + \sum_{s \in \Z} |C_{f,g}(s)|^2= \sum_{s\in \Zmz} C_{f,f}(s) \conj{C_{g,g}(s)}.
\end{equation}
Now the Cauchy--Schwarz inequality tells us that
\begin{equation}\label{Katherine}
|\Delta| \leq \sqrt{\left(\sum_{s \in \Zmz} |C_{f,f}(s)|^2\right) \left(\sum_{s \in \Zmz} |C_{g,g}(s)|^2\right)}.
\end{equation}
Use the first equality in \eqref{Irene} to substitute for $\Delta$ and divide by $C_{f,f}(0) C_{g,g}(0)$ (which is positive real since $f,g\not=0$), to obtain the equivalent inequality
\begin{equation}\label{Jane}
|-1+\CDF(f,g)| \leq \sqrt{\ADF(f) \ADF(g)},
\end{equation}
which is the bound we were to show.

The bound in the Cauchy--Schwarz inequality \eqref{Katherine} and in the equivalent result \eqref{Jane} is met if and only if there is some nonzero $\mu \in \C$ such that $C_{f,f}(s) = \mu C_{g,g}(s)$ for every $s \in \Zmz$.
(In principle, it would also be met if $C_{f,f}(s)=0$ for all $s \in \Zmz$ or if $C_{g,g}(s)=0$ for all $s\in \Zmz$, but these are impossible by \cref{Samuel}\eqref{Zosimos} because $\len f, \len g > 1$.)
If we have such a $\mu$, then we have
\begin{equation}\label{Elizabeth}
\Delta=\sum_{s \in \Zmz} C_{f,f}(s) \conj{C_{g,g}(s)} = \mu \sum_{s\in \Zmz} |C_{g,g}(s)|^2,
\end{equation}
and since $\Delta$ was shown to be real, this forces $\mu$ to be a real number.

If we have equality in \eqref{Jane} and $\mu$ is positive (resp., negative), then \eqref{Elizabeth} shows that $\Delta$ is positive (resp., negative), so that $\Delta/(C_{f,f}(0) C_{g,g}(0))=-1+\CDF(f,g)$ is positive (resp., negative), and so $-1+\CDF(f,g)=\sqrt{\ADF(f)\ADF(g)}$ and we meet the upper bound (resp., $-1+\CDF(f,g)=-\sqrt{\ADF(f)\ADF(g)}$ and we meet the lower bound).
These two extremes cannot be achieved simultaneously, as $\ADF(f)\ADF(g)\not=0$ because $\len f > 1$ and $\len g> 1$ (see \cref{Eugene}).
So, when $\min\{\len f,\len g\} > 1$, the necessary and sufficient condition for achieving the lower bound $-1+\CDF(f,g)=-\sqrt{\ADF(f)\ADF(g)}$ (i.e., $\PSC(f,g)=1$) is for there to be a negative real number $\mu$ such that $C_{f,f}(s)=\mu C_{g,g}(s)$ for all nonzero $s \in \Z$, which is to say, such that $(f,\sqrt{-\mu}g)$ is a Golay pair.
So to meet the lower bound, it is necessary that there be some $\lambda \in \C$ such that $(f,\lambda g)$ is a Golay pair.
Conversely, if there is a $\lambda \in \C$ such that $(f,\lambda g)$ is a Golay pair, then we know that $\lambda\not=0$ by \cref{Felix}, since $\len f > 1$.
Since \eqref{Eric} shows that scalar multiplication of a sequence by a unimodular complex number (e.g., $|\lambda|/\lambda$) does not change its autocorrelation values, $(f,|\lambda| g)=(f,\sqrt{-(-|\lambda|^2)} g)$ is also a Golay pair with $-|\lambda|^2$ a negative real number, so we meet the lower bound by what we have shown earlier in this paragraph.

For the rest of this proof, assume that $f$ and $g$ are unimodular with $\min\{\len f,\len g\}> 1$.
If $(f,g)$ is a Golay pair, then what we have already shown proves that our lower bound is met.
Conversely, if our lower bound is met, then we know that there is some nonzero $\lambda\in \C$ such that $(f,|\lambda| g)$ is a Golay pair.
\cref{Felix} shows that $\len f=\len(|\lambda| g)$, so that $\len f=\len g > 1$, and we can use \cref{Samuel}\eqref{Zosimos} to show that $C_{f,f}(\len f-1)+C_{|\lambda| g,|\lambda| g}(\len f-1)=f_{\deg f}\conj{f_{\ord f}}+|\lambda|^2 g_{\deg g} \conj{g_{\ord g}}$, which must equal $0$ by the Golay condition.
The various coefficients of $f$ and $g$ that appear in the last expression are all unimodular, so this forces $|\lambda|=1$, and so $(f,g)$ is a Golay pair.
So we meet the lower bound precisely when $(f,g)$ is a Golay pair, in which case \cref{Felicia} shows that $\len f=\len g$ and $\ADF(f)=\ADF(g)$. \hfill \qedsymbol

\begin{remark}
Theorem \ref{Gabriel} is reminiscent of a result of Liu and Guan \cite[Theorem 1]{Liu-Guan}.  Liu and Guan show that binary Golay pairs minimize a criterion that is different from the Pursley--Sarwate criterion in that they do not use the geometric mean of autocorrelation demerit factors, but (if their results are translated into the language of this paper) the arithmetic mean.  
Liu and Guan also consider the generalization to families that can have more than two sequences, and they generalize demerit factors to weighted versions.
For the remainder of this remark, we restrict attention to standard unweighted demerit factors of pairs (as in this paper) of binary sequences of the same length (as in Liu and Guan's paper), so we can compare their results with ours when both results are specialized to those situations where both may be invoked.  We translate their results into the notation of this paper: if $(f,g)$ is a pair of binary sequences of equal length, then Liu and Guan's inequality (11) becomes (after a significant amount of calculation)
\[
\frac{1}{4}\left(\CDF(f,f)+\CDF(f,g)+\CDF(g,f)+\CDF(g,g)\right) \geq 1,
\]
or equivalently, since $\CDF(f,f)=\ADF(f)+1$, $\CDF(g,g)=\ADF(g)+1$, and $\CDF(g,f)=\CDF(f,g)$, 
\begin{equation}\label{Edna}
\frac{\ADF(f)+\ADF(g)}{2} + \CDF(f,g) \geq 1.
\end{equation}
Liu and Guan's Theorem 1 asserts that this inequality becomes an exact equality if and only if $\{f,g\}$ is a complementary set.
This result is similar to \cref{Gabriel}, but we note that Liu and Guan's inequality has the arithmetic mean of the autocorrelation demerit factors while \cref{Gabriel} has the geometric mean.
\cref{Gabriel} immediately implies Liu and Guan's result by the arithmetic--geometric mean inequality, but Liu and Guan's result does not immediately imply \cref{Gabriel}.
\end{remark}

\section{Transformations of Golay Pairs}\label{Leonard}

In this section, we summarize known transformations for Golay pairs and show how they influence the autocorrelation and crosscorrelation behavior of the pairs.
In \cite[p.~496]{Golay-51} and \cite[p.~83]{Golay-61}, Golay lists six transformations, each of which takes a binary Golay pair $(f,g)$ to another binary Golay pair; these transformations are interchanging the two sequences, reversing the order of the terms in first sequence, reversing the order of the terms in the second sequence, negating the first sequence, negating the second sequence, and negating every other element in both sequences.
To represent these transformations in our Laurent polynomial formalism, if $f(z) \in \laur$, then we define the {\it conjugate reverse of $f(z)$}, written $f^\ddag(z)$, to be $f^\ddag(z)=z^{\ord f+\deg f} \conj{f(z)}$.
Recall our rule that $\infty+(-\infty)=0$, so that $0^\ddag=0$.
Conjugate reversal preserves degree and order ($\deg f^\ddag=\deg f$ and $\ord f^\ddag=\ord f$), is involutory ($f^{\ddag\ddag}=f$), and $|f^\ddag|^2=|z^{\ord f+\deg f} \conj{f}|^2=|f|^2$ for every $f \in \laur$.
If $m,n$ are integers with $m \leq n$ and $f(z)=f_m z^m + f_{m+1} z^{m+1} + \cdots + f_n z^n \in \laur$ with $f_m,f_n\not=0$, then $f^\ddag(z)=\conj{f_n} z^m + \conj{f_{n-1}} z^{m+1} +\cdots+\conj{f_m} z^n$, and so conjugate reverse of a contiguous sequence has the same support as the original sequence, but the coefficients are conjugated and arranged in reverse order.
We believe that this is the most useful generalization to complex sequences of Golay's transformation that reverses the order of a binary sequence.
Borwein and Mossinghoff \cite[p.~1159]{Borwein-Mossinghoff} use the reciprocal polynomial $f^*(z)=z^{\deg f} f(1/z)$ to reverse a polynomial $f(z)$ representing a binary sequence, and Katz, Lee, and Trunov (see \cite[p.~514]{Katz-Lee-Trunov-a} and \cite[p.~7728]{Katz-Lee-Trunov-b}) use the conjugate reciprocal polynomial $f^\dag(z)$ (which is obtained from $f^*(z)$ by conjugating every coefficient) as the generalization for polynomials with complex coefficients, but neither of these operations is invertible when one allows sequences whose constant coefficients equal zero, e.g., both $z$ and $z^2$ have reciprocal (and conjugate reciprocal) equal to $1$.
Even if one's main interest is in Golay pairs formed from binary sequences represented by polynomials with nonzero constant coefficients, the most comprehensive construction of such Golay pairs, due to Borwein and Ferguson \cite[Sections 4--5]{Borwein-Ferguson}, involves (in intermediate steps) sequences that can have $0$ for their constant coefficients.
Since we require groups of transformations that work on sequences such as these, the conjugate reversal operation defined here can be used, while the reciprocal and conjugate reciprocal cannot.

If $r(z) \in \laur$, we define the transformation $\subs_r \colon \laur^2\to\laur^2$ by $\subs_r(f(z),g(z))=(f(r(z)),g(r(z)))$, that is, $\subs_r$ substitutes $r(z)$ for the indeterminate $z$.
We can now define generalizations of Golay's six original transformations.
\begin{definition}[Elementary Golay transformations]
The {\it elementary Golay transformations} are the following transformations that map $\laur^2$ to itself.
\begin{enumerate}
\item $\swap(f,g) = (g,f)$ (swap the sequences in a pair),
\item $\conjug(f,g)=(\conj f,g)$ (conjugate the first sequence),
\item $\conjug'(f,g)=(f, \conj g)$ (conjugate the second sequence),
\item $\crev(f(z),g(z)) = (f^\ddag, g)$ (conjugate reverse the first sequence),
\item $\crev'(f(z),g(z)) = (f, g^\ddag)$ (conjugate reverse the second sequence),
\item $\srev(f(z),g(z))=(z^{\ord f+\deg f} f(z^{-1}),z^{\ord g+\deg g} g(z^{-1}))$ (simultaneously reverse the sequences),
\item $\scal_{p,q}(f,g)) = (p f,q g)$ (scale by any monomials $p$ and $q$ with $|p|^2=|q|^2$), and
\item $\subs_r(f(z),g(z)) = (f(r(z)), g(r(z)))$ (substitute a monomial $r(z)=w z^d$ with $|w|=1$ and $d\in\{-1,1\}$ for the indeterminate $z$).
\end{enumerate}
\end{definition}
Golay's six transformations are $\swap$, $\crev$, $\crev'$, $\scal_{-1,1}$, $\scal_{1,-1}$, and $\subs_{-z}$.
\begin{lemma}\label{Gertrude}
Each elementary Golay transformation is a permutation of $\laur^2$ whose inverse is another elementary Golay transformation: $\swap$, $\conjug$, $\conjug'$, $\crev$, $\crev'$, and $\srev$ are involutions, the inverse of $\scal_{p,q}$ is $\scal_{1/p,1/q}$, and the inverse of $\subs_{w z^d}$ is $\subs_{w^{-d} z^d}$.
\end{lemma}
\begin{proof}
All of the claims are easy to check, and note that when $p(z)$ and $q(z)$ are monomials with $|p(z)|^2=|q(z)|^2$, then $1/p(z)$ and $1/q(z)$ are also monomials and have $|1/p(z)|^2=|1/q(z)|^2$, and if $r(z)=w z^d$ is a monomial with $|w|=1$ and $d\in \{-1,1\}$, then $s(z)=w^{-d} z^d$ is a monomial of the same degree with $|w^{-d}|=1$.
\end{proof}
\begin{definition}[Golay group, $\Gol$]
The {\it Golay group}, written $\Gol$, is the group of permutations of $\laur^2$ generated by the elementary Golay transformations.
\end{definition}
We also define another type of transformation of $\laur^2$ that is not, in general, invertible.
\begin{definition}[Dilation]
For a nonzero integer $d$, the {\it dilation by $d$} is the transformation $\subs_{z^d}$.
\end{definition}
\begin{definition}[Extended Golay monoid, $\EGol$]
The {\it extended Golay mon-oid}, written $\EGol$, is the monoid of maps from $\laur^2$ to itself generated by the Golay group $\Gol$ and the dilation maps $\{\subs_{z^d}: d \in \Z, d\not=0\}$.
\end{definition}
We now examine sets of generators for $\Gol$ and $\EGol$.
\begin{lemma}\label{Manfred}
Let $S$ be the set whose elements are $\swap$, $\conjug$, $\crev$, $\srev$, $\scal_{p,q}$ for all monomials $p$ and $q$ with $|p|^2=|q|^2$, and $\subs_r$ for all monomials $r=w z^d$ with $|w|=1$ and $d\in\{-1,1\}$.
Then the monoid generated by $S$ is $\Gol$, and therefore the group generated by $S$ is $\Gol$.
Let $T$ be the set whose elements are $\swap$, $\conjug$, $\crev$, $\srev$, $\scal_{p,q}$ for all monomials $p$ and $q$ with $|p|^2=|q|^2$, and $\subs_r$ for all monomials $r=w z^d$ with $|w|=1$ and $d\not=0$.
Then the monoid generated by $T$ is $\EGol$.
\end{lemma}
\begin{proof}
Note that $S$ is the set obtained by removing $\conjug'$ and $\crev'$ from the set of elementary Golay transformations, but since $\conjug'=\swap\circ\conjug\circ\swap$ and $\crev'=\swap\circ\crev\circ\swap$, the monoid generated by the $S$ includes all the elementary Golay transformations, and since the set of elementary Golay transformations is closed under inversion by \cref{Gertrude}, the monoid generated by the $S$ is $\Gol$.
Every transformation of the form $\subs_r$ such that $r(z)=w z^d$ with $|w|=1$ and $d\not=0$ is in $\EGol$, since $\subs_r=\subs_{z^d} \circ \subs_{w z}$, which is a composition of a dilation and an element of $\Gol$, and so $T$ is a superset of $S$ but a subset of $\EGol$, so it generates a monoid including $\Gol$ and included in $\EGol$.
The set $T$ also contains all the dilations, so the monoid it generates must be all of $\EGol$.
\end{proof}
Golay was interested in his six original transformations because they map Golay pairs to Golay pairs.
We shall see that the same is true of elements of our extended Golay monoid, after recording without proof some very straightforward principles.
\begin{lemma}\label{Theresa}
Let $(a,b)\in\laur^2$ and $\gamma\in\Gol$, and set $(f,g)=\gamma(a,b)$.
If $\gamma=\swap$, then $(|a|^2,|b|^2)=(|g|^2,|f|^2)$.
If $\gamma\in\{\conjug,\conjug',\crev,\crev'\}$, then $(|f|^2,|g|^2)=(|a|^2,|b|^2)$.
If $\gamma=\scal_{p,q}$ where $p$ and $q$ are monomials with $|p|^2=|q|^2$, then $|p|^2$ is a positive real number and $(|f|^2,|g|^2)=(|p|^2 |a|^2,|p|^2 |b|^2)$.
If $\gamma=\srev$, then $|f|^2_k=|a|^2_{-k}$ and $|g|^2_k=|b|^2_{-k}$ for every $k \in\Z$.
If $\gamma=\subs_r$ where $r=w z^d$ with $|w|=1$ and $d\not=0$, then $|f|^2_j=|g|^2_j=0$ for all $j \in \Z$ with $d\nmid j$, and $|f|^2_{d k}=w^k |a|^2_k$ and $|g|^2_{d k}=w^k |b|^2_k$ for every $k \in \Z$.
\end{lemma}
Now we prove that transformations from $\EGol$ preserve Golay complementarity.
\begin{lemma}\label{Zachary}
If $(a,b) \in \laur^2$ and $\gamma \in \EGol$, then $(a,b)$ is a Golay pair if and only if $\gamma(a,b)$ is a Golay pair.
\end{lemma}
\begin{proof}
In view of \cref{Manfred}, it suffices to show this in the case where $\gamma$ is one of $\swap$, $\conjug$, $\crev$, $\srev$, $\scal_{p,q}$ (for any monomials $p,q$ with $|p|^2=|q|^2$), $\subs_r$ (for any nonconstant monomial $r$ with a unimodular coefficient).
If $\gamma$ is $\swap$, $\conjug$, $\crev$, or $\scal_{p,q}$, and we let $(f,g)=\gamma(a,b)$, then \cref{Theresa} shows that there is some positive real number $\lambda$ such that $\{|f|^2,|g|^2\}=\{\lambda |a|^2,\lambda |b|^2\}$, and so $|f|^2+|g|^2=\lambda(|a|^2+|b|^2)$ is constant if and only if $|a|^2+|b|^2$ is constant.
If $\gamma=\srev$, then \cref{Theresa} shows that $(|f|^2+|g|^2)_k=(|a|^2+|b|^2)_{-k}$ for every $k\in\Z$, so $|f|^2+|g|^2$ is constant if and only if $|a|^2+|b|^2$ is constant.
If $\gamma=\subs_r$ for $r(z)=w z^d$ with $|w|=1$ and $d\not=0$, then \cref{Theresa} shows that $|f|^2+|g|^2$ only has terms whose degrees are multiples of $d$, and $(|f|^2+|g|^2)_{d k}=w^k (|a|^2+|b|^2)_k$ for every $k\in\Z$, so that $|f|^2+|g|^2$ is constant if and only if $|a|^2+|b|^2$ is constant.
\end{proof}
Our transformations also preserve the Pursley--Sarwate criterion.
\begin{lemma}\label{Cameron}
Let $(a,b) \in \laur^2$ with $a,b\not=0$, let $\gamma \in \EGol$, and set $(f,g)=\gamma(a,b)$.
Then $(f,g)$ is isoenergetic if and only if $(a,b)$ is isoenergetic.
Also $\{\ADF(f),\ADF(g)\}=\{\ADF(a),\ADF(b)\}$ and $\CDF(f,g) = \CDF(a,b)$, and thus $\PSC(f,g)=\PSC(a,b)$.
\end{lemma}
\begin{proof}
In view of \cref{Manfred}, it suffices to show this in the case where $\gamma$ is one of $\swap$, $\conjug$, $\crev$, $\srev$, $\scal_{p,q}$ (for any monomials $p,q$ with $|p|^2=|q|^2$), or $\subs_r$ (for any nonconstant monomial $r$ with a unimodular coefficient).
If $\gamma$ is $\swap$, $\conjug$, $\crev$, or $\scal_{p,q}$, then \cref{Theresa} shows that there is some positive real number $\lambda$ such that $\{|f|^2,|g|^2\}=\{\lambda |a|^2,\lambda |b|^2\}$.
Thus $|f|^2_0=|g|^2_0$ if and only if $|a|^2_0=|b|^2_0$, so by \eqref{Dorothy} we see that $(f,g)$ is isoenergetic if and only if $(a,b)$ is.
Also $\{|f|^4_0/(|f|^2_0)^2,|g|^4_0/(|g|^2_0)^2\}=\{|a|^4_0/(|a|^2_0)^2,|b|^4_0/(|b|^2_0)^2\}$, and so by \eqref{Vera} we have $\{\ADF(f),\ADF(g)\}=\{\ADF(a),\ADF(b)\}$.
Similarly, $|f g|^2_0/(|f|^2_0 |g|^2_0)=|a b|^2_0/(|a|^2_0 |b|^2_0)$, and so by \eqref{Bart} we have $\CDF(a,b)=\CDF(f,g)$.

Now assume either that $\gamma=\srev$, in which case \[(f(z),g(z))=(z^{\ord a+\deg a}a(z^{-1}),z^{\ord b+\deg b} b(z^{-1})),\] or that $\gamma=\subs_r$, in which case $(f(z),g(z))=(a(r(z)),b(r(z)))$.
In either case $(|f(z)|^2,|g(z)|^2)=(|a(s(z))|^2,|b(s(z))|^2)$ for some nonconstant monomial $s(z)$ with a unimodular coefficient.
Thus, for every $h(z) \in \laur$, the fact that $s(z)$ is a nonconstant monomial with a unimodular coefficient makes $|h(s(z))|^2_0=|h|^2_0$.
If we consider $h=a$ (resp., $h=b$) and look at \eqref{Dorothy}, we see that $\normt{a}=\normt{f}$ (resp., $\normt{b}=\normt{g}$), and so $(f,g)$ is isoenergetic if and only if $(a,b)$ is isoenergetic.
On the other hand, if we consider $h\in\{a^2,a\}$ (resp., $h\in\{b^2,b\}$) and look at \eqref{Vera}, we see that $\ADF(f)=\ADF(a)$ (resp., $\ADF(g)=\ADF(b)$), and if we consider $h\in\{a b,a,b\}$ and look at \eqref{Bart}, we see that $\CDF(f,g)=\CDF(a,b)$.
\end{proof}
Sometimes we are only interested in the transformations that preserve certain properties of pairs, so we define some special subgroups of $\Gol$.
\begin{definition}[Stationary Golay Group, $\SGol$]
The {\it stationary Golay group}, written $\SGol$, is the subgroup of $\Gol$ generated by $\swap$, $\crev$, $\crev'$, $\srev$, all transformations $\scal_{u,v}$ where $u,v$ are nonzero complex numbers with $|u|=|v|$, and all transformations $\subs_{w z}$, where $w$ is a unimodular complex number.
\end{definition}
\begin{remark}\label{Samantha}
Suppose that $(a,b)\in\laur^2$, $\gamma \in \SGol$, and $(f,g)=\gamma(a,b)$.
Then one easily sees that
\[
\{(\ord f,\deg f),(\ord g,\deg g)\}=\{(\ord a,\deg a),(\ord b,\deg b)\}
\]
since all the generators and inverses of generators of $\SGol$ except for $\swap$ preserve the orders and degrees of both sequences in the pair, while $\swap$ swaps the two sequences and the pair.
In particular, transformations in $\SGol$ preserve the minimum and maximum values of order (or of degree, or of length) for the two sequences in the pair.
Therefore, if $\ord a=\ord b$ (resp., $\deg a=\deg b$, $\len a=\len b$), then $\ord(f,g)=\ord(a,b)$ (resp., $\deg(f,g)=\deg(a,b)$, $\len(f,g)=\len(a,b)$).
\end{remark}
\begin{definition}[Unimodular Golay Group $\UGol$]
The {\it unimodular Golay group}, written $\UGol$, is the subgroup of $\Gol$ generated by $\swap$, $\crev$, $\crev'$, $\srev$, all transformations $\scal_{u,v}$ where $u,v$ are unimodular complex numbers, and all transformations $\subs_{w z}$ where $w$ is a unimodular complex number.
\end{definition}
\begin{remark}\label{Ulysses}
If $(f,g) \in \laur$ and $\gamma\in \UGol$, then one readily sees $(f,g)$ is unimodular if and only if $\gamma(f,g)$ is unimodular.
Also, $\UGol$ is clearly a subgroup of $\SGol$, and so it has the properties mentioned in \cref{Samantha}.
\end{remark}
\begin{definition}[Binary Golay Group $\BGol$]
The {\it binary Golay group}, written $\BGol$, is the subgroup of $\Gol$ generated by $\swap$, $\crev$, $\crev'$, $\srev$, $\scal_{-1,1}$, $\scal_{1,-1}$, and $\subs_{-z}$.
\end{definition}
\begin{remark}\label{Barbara}
The binary Golay group is simply the group generated by Golay's six transformations along with $\srev$, but one should note that if we are applying transformations in $\BGol$ to sequences with real terms, then $\srev$ has the same effect as $\crev \circ \crev'$.
Clearly $\BGol$ is a subgroup of $\UGol$, since it is generated by a subset of $\UGol$'s generators.
If $(f,g) \in \laur$ and $\gamma\in \BGol$, then one readily sees $(f,g)$ is binary if and only if $\gamma(f,g)$ is binary.
Also, $\BGol$ is clearly a subgroup both of $\SGol$ and of $\UGol$, and so it has the properties mentioned in Remarks \ref{Samantha} and \ref{Ulysses}.
\end{remark}

\section{Constructions of Golay Pairs}\label{Clarence}

In this section we describe construction methods, each of which takes two Golay pairs, $(a,b)$ and $(c,d)$, and produces another Golay pair $(f,g)$.
Golay, Turyn, and Borwein--Ferguson gave various methods of this kind.
We devise a single simple method which, using the transformations from the extended Golay monoid and the pair $(1,1)$ (which is obviously Golay), can construct all Golay pairs obtainable by these earlier methods.
\begin{construction}[Weaving]
Let $a,b,c,d \in \laur$.
Then the {\it weave of $(a,b)$ with $(c,d)$}, denoted $(a,b)\ltimes (c,d)$ or $(c,d)\rtimes (a,b)$, is the pair $(f,g)$, where
\begin{align*}
f(z) & = a(z)c(z) + b(z)d(z)\\
g(z) & = a(z)\conj{d(z)} - b(z)\conj{c(z)}.
\end{align*}
\end{construction}
We want to show that this construction produces Golay pairs when the inputs are Golay pairs.
\begin{lemma}\label{Hugo}
Let $a,b,c,d \in \laur$ and let $(f,g)=(a,b)\ltimes(c,d)$.
Then
\begin{align*}
|f|^2 & = |a c|^2 + |b d|^2 + 2 \Re(a c \conj{b d}) \\
|g|^2 & = |a d|^2 + |b c|^2 - 2 \Re(a c \conj{b d}), \text{ and} \\
|f|^2 + |g|^2 & = (|a|^2 + |b|^2)(|c|^2 + |d|^2).
\end{align*}
\end{lemma}
\begin{proof}
Since $f=a c + b d$, we have $|f|^2=|a c|^2 + |b d|^2 + a c \conj{b d} + \conj{a c} b d$, which verifies the first formula.
And since $g=a \conj{d} - b \conj{c}$, we have $|g|^2=|a d|^2 + |b c|^2 - a c \conj{b d} - \conj{a c} b d$, which verifies the second formula.
And then
\begin{align*}
|f|^2 + |g|^2 & = |a c|^2 + |b d|^2 + 2\Re(a c\conj{b d}) + |a d|^2 + |b c|^2 -2 \Re(a c \conj{b d}) \\
& = (|a|^2 + |b|^2)(|c|^2 + |d|^2). \qedhere
\end{align*}
\end{proof}
\begin{corollary}\label{Ursula}
Let $a,b,c,d \in \laur$.  If both $(a,b)$ and $(c,d)$ are not equal to $(0,0)$, then $(a,b)\ltimes(c,d)$ is a Golay pair if and only if both $(a,b)$ and $(c,d)$ are Golay pairs.
If $(a,b)=(0,0)$ or $(c,d)=(0,0)$, then $(a,b)\ltimes(c,d)=(0,0)$, which is a Golay pair.
\end{corollary}
\begin{proof}
The claim when $(a,b)$ or $(c,d)$ is $(0,0)$ is clear, so suppose neither is $(0,0)$, and let $(f,g)=(a,b)\ltimes(c,d)$.
\cref{Hugo} shows that $|f|^2+|g|^2$ is constant if both $|a|^2+|b|^2$ and $|c|^2+|d|^2$ are.
By \eqref{Dorothy} and the fact that $(a,b),(c,d)\not=(0,0)$, we see that $|a|^2+|b|^2$ and $|c|^2+|d|^2$ have nonzero constant terms.
If either of them were nonconstant, then their product $|f|^2+|g|^2$ would have more than one monomial, and hence be nonconstant.
So $|f|^2+|g|^2$ is a constant if and only if both $|a|^2+|b|^2$ and $|c|^2+|d|^2$ are, so the result follows from \cref{Allison}.
\end{proof}
The following technical lemma, which will be used later, considers how the weaving construction influences expressions that appear in our Laurent polynomial formulae \eqref{Vera} and \eqref{Bart} for autocorrelation and crosscorrelation demerit factors.
\begin{lemma}\label{Roger}
If $a,b,c,d \in \laur$ and $(f,g)=(a,b)\ltimes(c,d)$, then we have
\begin{align*}
|f|^4 & = |a c|^4 + |b d|^4 + 4|a b c d|^2  +4(|a c|^2 + |b d|^2)\Re(a c\conj{b d}) + 2\Re((a c\conj{b d})^2), \\
|g|^4 & = |a d|^4 + |b c|^4 + 4|a b c d|^2 -4(|a d|^2 + |b c|^2)\Re(a c\conj{b d}) + 2\Re((a c\conj{b d})^2),
\end{align*}
and
\begin{multline*}
|fg|^2  = |a b|^2 (|c|^4 + |d|^4) + |c d|^2(|a|^4 + |b|^4) - 2 |a b c d|^2 \\
- 2(|a|^2 - |b|^2)(|c|^2 - |d|^2)\Re(a c\conj{b d}) -2\Re((a c\conj{b d})^2).
\end{multline*}
\end{lemma}
\begin{proof}
By \cref{Hugo}, we have $|f|^2=|a c|^2+|b d|^2+2 \Re(a c \conj{b d})$.
Square this and use the identity $4 \Re(u)^2 = 2 \Re(u^2) + 2 |u|^2$ to obtain the result for $|f|^4$.
One obtains the result for $|g|^4$ the same way by replacing $c$ and $d$ with $\conj{d}$ and $-\conj{c}$, respectively.
From our expressions for $|f|^4$ and $|g|^4$, we have
\begin{multline*}
|f|^4+|g|^4 =(|a|^4+|b|^4)(|c|^4+|d|^4) + 8|a b c d|^2 \\
+4(|a|-|b|^2)(|c|^2-|d|^2) \Re(a c\conj{b d}) + 4\Re((a c\conj{b d})^2).
\end{multline*}
Then note that
\[
|f g|^2 = \frac{(|f|^2+|g|^2)^2-(|f|^4+|g|^4)}{2},
\]
and then we use the expression for $|f|^2+|g|^2$ from \cref{Hugo} and our expression for $|f|^4+|g|^4$ to deduce the expression for $|f g|^2$.
\end{proof}
Now we describe the historical methods that were used to construct binary Golay pairs.
We present them in slightly more general forms than the originals so as to make them also suitable for constructing Golay pairs with complex terms. 
The first two constructions are due to Golay: see \cite[eqs.~(10),(11)]{Golay-61}.
\begin{construction}[Golay concatenation]\label{GCC}
If $(a,b)$ and $(c,d)$ are Golay pairs with $\ord c+\deg c=\ord d+\deg d$, and $m$ and $n$ are integers with $m\not=0$, and we let $f(z) = a(z) c(z^m) + z^{m n} b(z) d(z^m)$ and $g(z) = a(z) d^\ddag(z^m) - z^{m n} b(z) c^\ddag(z^m)$, then
\[
(f,g)=\scal_{1,z^{m(\ord c+\deg c)}}\left((\scal_{1,z^{m n}}(a,b))\ltimes(\subs_{z^m}(c,d))\right),
\]
which is a Golay pair.
In particular, if $(a,b)$ and $(c,d)$ are unimodular (resp., binary) Golay pairs with $\len(a,b)=m$, $\len(c,d)=n$, $\ord a=\ord b$, and $\ord c=\ord d$, then $(f,g)$ is a unimodular (resp., binary) Golay pair with $\len(f,g)=2 m n$.
\end{construction}
\begin{proof}
It is not difficult to check the formula for $(f,g)$, from which it follows that $(f,g)$ is Golay via \cref{Zachary} and \cref{Ursula}.
The second claim is clear if $(a,b)$ or $(c,d)$ is $(0,0)$, and otherwise follows from the fact that the coefficient of $z^j$ in $f(z)$ (resp., $g(z)$) is $0$ unless $\ord a+ m \ord c \leq j \leq m n+\deg a+ m \deg c$, in which case it is the either the product of a nonzero coefficient of $a$ with one of $c$ or the product of a nonzero coefficient of $b$ with one of $d$ (resp., either the product of a nonzero coefficient of $a$ with the conjugate of one of $d$ or the product of a nonzero coefficient of $b$ with the conjugate of one of $c$).
\end{proof}
\begin{construction}[Golay interleaving \cite{Golay-61}]\label{GIC}
If $(a,b)$ and $(c,d)$ are Golay pairs with $\ord c+\deg c=\ord d+\deg d$, and $m$ is a nonzero integer, and we let $f(z) = a(z) c(z^{2 m}) + z^m b(z) d(z^{2 m})$ and $g(z) = a(z) d^\ddag(z^{2 m}) - z^m b(z) c^\ddag(z^{ 2 m})$, then
\[
(f,g)=\scal_{1,z^{2 m(\ord c+\deg c)}}\left((\scal_{1,z^m}(a,b))\ltimes(\subs_{z^{2 m}}(c,d))\right),
\]
which is a Golay pair.
In particular, if $(a,b)$ and $(c,d)$ are unimodular (resp., binary) Golay pairs with $\len(a,b)=m$, $\len(c,d)=n$, $\ord a=\ord b$, and $\ord c=\ord d$, then $(f,g)$ is a unimodular (resp., binary) Golay pair with $\len(f,g)=2 m n$.
\end{construction}
\begin{proof}
It is not difficult to check the formula for $(f,g)$, from which it follows that $(f,g)$ is Golay via \cref{Zachary} and \cref{Ursula}.
The second claim is clear if $(a,b)$ or $(c,d)$ is $(0,0)$, and otherwise follows from the fact that the coefficient of $z^j$ in $f(z)$ (resp., $g(z)$) is $0$ unless $\ord a+2 m\ord c \leq j \leq m+\deg a+2 m\deg c$, in which case it is the either the product of a nonzero coefficient of $a$ with one of $c$ or the product of a nonzero coefficient of $b$ with one of $d$ (resp., either the product of a nonzero coefficient of $a$ with the conjugate of one of $d$ or the product of a nonzero coefficient of $b$ with the conjugate of one of $c$).
\end{proof}
The next construction was presented by Borwein and Ferguson \cite[p.~975]{Borwein-Ferguson} as being Golay's interleaving construction.
In fact, it is a slightly different (but still valid) way of combining two Golay pairs to get a longer one.
\begin{construction}[Borwein-Ferguson interleaving \cite{Borwein-Ferguson}]\label{BFIC}
If $(a,b)$ and $(c,d)$ are Golay pairs with $\ord c+\deg c=\ord d+\deg d$ and $m$ is a nonzero integer, and we let $f(z) = a(z^2) c(z^{2 m}) + z b(z^2) d(z^{2 m})$ and let $g(z) = a(z^2) d^\ddag(z^{2 m}) - z b(z^2) c^\ddag(z^{2 m})$, then
\[
(f,g)=\scal_{1,z^{2 m(\ord c+\deg c)}}\left((\scal_{1,z} \subs_{z^2}(a,b))\ltimes(\subs_{z^{2 m}}(c,d))\right),
\]
which is a Golay pair.
In particular, if $(a,b)$ and $(c,d)$ are unimodular (resp., binary) Golay pairs with $\len(a,b)=m$, $\len(c,d)=n$, $\ord a=\ord b$, and $\ord c=\ord d$, then $(f,g)$ is a unimodular (resp., binary) Golay pair with $\len(f,g)=2 m n$.
\end{construction}
\begin{proof}
It is not difficult to check the formula for $(f,g)$, from which it follows that $(f,g)$ is Golay via \cref{Zachary} and \cref{Ursula}.
The second claim is clear if $(a,b)$ or $(c,d)$ is $(0,0)$, and otherwise follows from the fact that the coefficient of $z^j$ in $f(z)$ (resp., $g(z)$) is $0$ unless $2 \ord a+2 m \ord c \leq 1+2 \deg a + 2 m \deg c$, in which case it is the either the product of a nonzero coefficient of $a$ with one of $c$ or the product of a nonzero coefficient of $b$ with one of $d$ (resp., either the product of a nonzero coefficient of $a$ with the conjugate of one of $d$ or the product of a nonzero coefficient of $b$ with the conjugate of one of $c$).
\end{proof}
The final construction is due to Turyn \cite[Lemma 5]{Turyn}.\footnote{Note that Turyn uses the notation $A\times B$ to denote the tensor product $B\otimes A$; this is clear when one compares his summaries (immediately preceding Lemma 5) of Golay's constructions with the originals in \cite[eq.~(10),(11)]{Golay-61}.}
\begin{construction}[Turyn]
If $(a,b)$ and $(c,d)$ are Golay pairs with $\ord a+\deg a=\ord b+\deg b$ and $m$ is a nonzero integer, and we let
\begin{align*}
f(z) & = a(z) \left(\frac{c(z^m) + d(z^m)}{2}\right) - b^\ddag(z) \left(\frac{c(z^m) - d(z^m)}{2}\right) \text{ and}\\
g(z) & = b(z) \left(\frac{c(z^m) + d(z^m)}{2}\right) + a^\ddag(z) \left(\frac{c(z^m) - d(z^m)}{2}\right),
\end{align*}
then $(f,g)$ is \[\scal_{1/2,-z^{\ord a+\deg a}/2} \left(\left(\subs_{z^m}\left((c,d)\ltimes(1,1)\right)\right)\ltimes (\scal_{1,-1} \crev' (a,b))\right),\] which is a Golay pair.
In particular, if $(a,b)$ is a unimodular (resp., binary) Golay pair and $(c,d)$ is a binary Golay pair, and if $\len(a,b)=m$, $\len(c,d)=n$, $\ord a=\ord b$, and $\ord c=\ord d$, then $(f,g)$ is a unimodular (resp., binary) Golay pair with $\len(f,g)=m n$.
\end{construction}
\begin{proof}
It is not difficult to check the formula for $(f,g)$, from which it follows that $(f,g)$ is Golay via \cref{Zachary}, \cref{Ursula}, and the fact that $(1,1)$ is a Golay pair.
The second claim is clear if $(a,b)$ or $(c,d)$ is $(0,0)$.
Otherwise, we first note that $(c+d)/2$ and $(c-d)/2$ are sequences whose coefficients for $z^j$ both vanish unless $\ord c \leq j \leq \deg c$, in which case precisely one of them vanishes and the other is in $\{-1,1\}$.
Thus the coefficient of $z^j$ in $f(z)$ (resp., in $g(z)$) is $0$ unless $\ord a+m \ord c \leq j \leq \deg a+m\deg c$, in which case it is equal to plus or minus either a nonzero coefficient of $a$ or else the conjugate one of $b$ (resp., plus or minus either a nonzero coefficient of $b$ or else the conjugate of one of $a$).
\end{proof}
For arbitrary nonnegative integers $a$, $b$, and $c$, one can construct a binary Golay pair of length $2^a 10^b 26^c$ using Turyn's construction and the following sequence pairs (represented using $+$ to denote $+1$ and $-$ to denote $-1$).
\begin{itemize}
\item Length 1:
\begin{center}
\begin{tikzpicture}
\node at (0,0.5) {$+$};
\node at (0,0) {$+$};
\end{tikzpicture}
\end{center}
\item Length 2:
\begin{center}
\begin{tikzpicture}
\node at (0,0.5) {$+$};
\node at (0.5,0.5) {$+$};
\node at (0,0) {$+$};
\node at (0.5,0) {$-$};
\end{tikzpicture}
\end{center}
\item Length 10:
\begin{center}
\begin{tikzpicture}
\node at (0,0.5) {$+$};
\node at (0.4,0.5) {$+$};
\node at (0.8,0.5) {$+$};
\node at (1.2,0.5) {$+$};
\node at (1.6,0.5) {$+$};
\node at (2.0,0.5) {$-$};
\node at (2.4,0.5) {$+$};
\node at (2.8,0.5) {$-$};
\node at (3.2,0.5) {$-$};
\node at (3.6,0.5) {$+$};
\node at (0,0) {$+$};
\node at (0.4,0) {$+$};
\node at (0.8,0) {$-$};
\node at (1.2,0) {$-$};
\node at (1.6,0) {$+$};
\node at (2.0,0) {$+$};
\node at (2.4,0) {$+$};
\node at (2.8,0) {$-$};
\node at (3.2,0) {$+$};
\node at (3.6,0) {$-$};
\end{tikzpicture}
\end{center}
or
\begin{center}
\begin{tikzpicture}
\node at (0,0.5) {$+$};
\node at (0.4,0.5) {$+$};
\node at (0.8,0.5) {$-$};
\node at (1.2,0.5) {$+$};
\node at (1.6,0.5) {$+$};
\node at (2.0,0.5) {$+$};
\node at (2.4,0.5) {$+$};
\node at (2.8,0.5) {$+$};
\node at (3.2,0.5) {$-$};
\node at (3.6,0.5) {$-$};
\node at (0.0,0) {$+$};
\node at (0.4,0) {$+$};
\node at (0.8,0) {$-$};
\node at (1.2,0) {$+$};
\node at (1.6,0) {$-$};
\node at (2.0,0) {$+$};
\node at (2.4,0) {$-$};
\node at (2.8,0) {$-$};
\node at (3.2,0) {$+$};
\node at (3.6,0) {$+$};
\end{tikzpicture}
\end{center}
\item Length 26:
\begin{center}
\begin{tikzpicture}
\node at (0,0.5) {$+$};
\node at (0.4,0.5) {$+$};
\node at (0.8,0.5) {$+$};
\node at (1.2,0.5) {$+$};
\node at (1.6,0.5) {$-$};
\node at (2.0,0.5) {$+$};
\node at (2.4,0.5) {$+$};
\node at (2.8,0.5) {$-$};
\node at (3.2,0.5) {$-$};
\node at (3.6,0.5) {$+$};
\node at (4.0,0.5) {$-$};
\node at (4.4,0.5) {$+$};
\node at (4.8,0.5) {$+$};
\node at (5.2,0.5) {$+$};
\node at (5.6,0.5) {$+$};
\node at (6.0,0.5) {$+$};
\node at (6.4,0.5) {$-$};
\node at (6.8,0.5) {$+$};
\node at (7.2,0.5) {$-$};
\node at (7.6,0.5) {$-$};
\node at (8.0,0.5) {$-$};
\node at (8.4,0.5) {$+$};
\node at (8.8,0.5) {$+$};
\node at (9.2,0.5) {$-$};
\node at (9.6,0.5) {$-$};
\node at (10.0,0.5) {$-$};
\node at (0,0) {$+$};
\node at (0.4,0) {$+$};
\node at (0.8,0) {$+$};
\node at (1.2,0) {$+$};
\node at (1.6,0) {$-$};
\node at (2.0,0) {$+$};
\node at (2.4,0) {$+$};
\node at (2.8,0) {$-$};
\node at (3.2,0) {$-$};
\node at (3.6,0) {$+$};
\node at (4.0,0) {$-$};
\node at (4.4,0) {$+$};
\node at (4.8,0) {$-$};
\node at (5.2,0) {$+$};
\node at (5.6,0) {$-$};
\node at (6.0,0) {$-$};
\node at (6.4,0) {$+$};
\node at (6.8,0) {$-$};
\node at (7.2,0) {$+$};
\node at (7.6,0) {$+$};
\node at (8.0,0) {$+$};
\node at (8.4,0) {$-$};
\node at (8.8,0) {$-$};
\node at (9.2,0) {$+$};
\node at (9.6,0) {$+$};
\node at (10.0,0) {$+$};
\end{tikzpicture}
\end{center}
\end{itemize}
We have listed two Golay pairs of length $10$ that are inequivalent modulo the action of the binary Golay group $\BGol$ (all other binary Golay pairs of length $10$ are equivalent to one of these); for the other lengths listed above there is only one binary Golay pair of that length modulo the action of $\BGol$ (see \cite[Table 1]{Djokovic}).
We note that each sequence in the second Golay pair of length $10$ above can be obtained from the corresponding sequence in the first Golay pair of length $10$ by selecting the next-to-leftmost term and then selecting every third term, proceeding cyclically (cf.~\cite[p.~86]{Golay-61}).

{\Dbar}okovi\'c \cite[Definition 2.1]{Djokovic} defines a {\it constructible} binary Golay pair to be one that is equivalent, modulo the action of the binary Golay group $\BGol$, to a binary Golay pair that can be generated from two smaller binary Golay pairs via a variant of Turyn's construction.
In addition to the binary Golay pairs of lengths $1$, $2$, $10$, and $26$ displayed above, {\Dbar}okovi\'c shows that (up to equivalence modulo $\BGol$) there are two non-constructible binary Golay pairs of length $16$, one of length $20$, and forty-four of length $32$.
Borwein and Ferguson \cite[pp.~975--978]{Borwein-Ferguson} devise a procedure that produces all binary Golay pairs of length less than $100$ starting from a set of only five starting binary Golay pairs.
Their procedure, based on Constructions \ref{GCC} and \ref{BFIC} and transformations from  $\BGol$, allows for non-binary Golay pairs whose terms are in $\{1,-1,0\}$ in intermediate steps, and their starting binary Golay pairs are the pairs of lengths $1$, $10$, and $26$ listed above, along with the following pair of length $20$.
\begin{itemize}
\item 
Length 20:
\begin{center}
\begin{tikzpicture}
\node at (0,0.5) {$+$};
\node at (0.4,0.5) {$+$};
\node at (0.8,0.5) {$+$};
\node at (1.2,0.5) {$+$};
\node at (1.6,0.5) {$-$};
\node at (2.0,0.5) {$+$};
\node at (2.4,0.5) {$+$};
\node at (2.8,0.5) {$+$};
\node at (3.2,0.5) {$+$};
\node at (3.6,0.5) {$+$};
\node at (4.0,0.5) {$-$};
\node at (4.4,0.5) {$-$};
\node at (4.8,0.5) {$-$};
\node at (5.2,0.5) {$+$};
\node at (5.6,0.5) {$-$};
\node at (6.0,0.5) {$+$};
\node at (6.4,0.5) {$-$};
\node at (6.8,0.5) {$+$};
\node at (7.2,0.5) {$+$};
\node at (7.6,0.5) {$-$};
\node at (0,0) {$+$};
\node at (0.4,0) {$+$};
\node at (0.8,0) {$+$};
\node at (1.2,0) {$+$};
\node at (1.6,0) {$-$};
\node at (2.0,0) {$+$};
\node at (2.4,0) {$-$};
\node at (2.8,0) {$-$};
\node at (3.2,0) {$-$};
\node at (3.6,0) {$+$};
\node at (4.0,0) {$+$};
\node at (4.4,0) {$-$};
\node at (4.8,0) {$-$};
\node at (5.2,0) {$+$};
\node at (5.6,0) {$+$};
\node at (6.0,0) {$-$};
\node at (6.4,0) {$+$};
\node at (6.8,0) {$-$};
\node at (7.2,0) {$-$};
\node at (7.6,0) {$+$};
\end{tikzpicture}
\end{center}
\end{itemize}
Since Borwein and Ferguson obtain every binary Golay pair of length less than $100$ starting from the specific Golay pairs of lengths $1$, $10$, $20$, and $26$ listed above by means of the transformations in $\BGol$ along with Constructions \ref{GCC} and \ref{BFIC}, the fact that these constructions can be obtained from the weaving construction and transformations in $\EGol$ shows that every binary Golay pair of length $100$ can be obtained from Borwein and Ferguson's five starting Golay pairs, the transformations in $\EGol$, and the weaving construction.

\section{Iterated Golay-Rudin-Shapiro Construction}\label{Taylor}

In this section we show how the demerit factors of sequences in Golay pairs change when we repeatedly apply a simple construction that produces longer and longer Golay pairs.
Before Golay presented \cref{GCC}, he had already devised \cite[p.~469]{Golay-51} the special case of it where the second pair used in the construction is $(c,d)=(1,1)$ and the parameter $n$ is $1$.
At almost the same time, Shapiro \cite[pp.~39--40]{Shapiro} devised the same special case of \cref{GCC} in his studies of Littlewood polynomials with small $L^\infty$ norm on the complex unit circle, and this was later rediscovered by Rudin \cite[eq.~(1.5)]{Rudin}; for this reason we call this special case of \cref{GCC} the Golay--Rudin--Shapiro construction.
\begin{construction}[Golay-Rudin-Shapiro]\label{GRS}
If $(a,b) \in \laur^2$ and $m$ is an integer, then we set 
\[
\GRS((a,b),m)=(\scal_{1,z^m}(a,b))\ltimes(1,1)=(a+z^m b,a-z^m b).
\]
We prove some properties of our construction.
\begin{enumerate}[(i).]
\item If $(a,b)$ is a Golay pair, then so is $\GRS((a,b),m)$.
\item If $\ord a=\ord b$ and $m>0$, then $\GRS((a,b),m)$ is a pair of order $\ord a$.
\item If $\len a=\len b=m$ and $\ord a=\ord b$, then $\GRS((a,b),m)$ is a pair of length $2 m$ and order $\ord a$.
\item If $(a,b)$ is a unimodular (resp., binary) pair with $\len a=\len b=m$ and $\ord a=\ord b$, then $\GRS((a,b),m)$ is a unimodular (resp., binary) pair of length $2 m$ and order $\ord a$.
\end{enumerate}
\end{construction}
\begin{proof}
The claim that $\GRS((a,b),m)$ is a Golay pair whenever $(a,b)$ is follows from  \cref{Zachary} and \cref{Ursula} and the fact that $(1,1)$ is a Golay pair, and the other claims are clear from the second expression for $\GRS((a,b),m)$.
\end{proof}
We now iterate the previous construction, also allowing for the application of transformations from $\EGol$.
\begin{construction}[Iterated Golay-Rudin-Shapiro]\label{IGRS}
Let $(a,b) \in \laur^2$, let $m \in \Z$, and let $\gamma=(\gamma_0,\gamma_1,\ldots)$ be a sequence of transformations from $\EGol$.
For $n \in \N$, we define $\GRS_\gamma^n((a,b),m)$ recursively by setting $\GRS_\gamma^0((a,b),m) =\gamma_0(a,b)$ and for $n > 0$ setting
\[
\GRS_\gamma^{n+1}=\gamma_{n+1}(\GRS(\GRS_\gamma^n((a,b),m),2^n m)).
\]
We prove some properties of our iterated construction.
\begin{enumerate}[(i).]
\item If $(a,b)$ is a Golay pair, then so is $\GRS_\gamma^n((a,b),m)$ for every $n \in \N$.
\item If $\ord a=\ord b$ and $m>0$, and $\gamma_j \in \SGol$ for every $j \in \N$, then $\GRS_\gamma^n((a,b),m)$ is a pair of order $\ord a$ for each $n\in\N$.
\item If $\len a=\len b=m$, $\ord a=\ord b$, and $\gamma_j \in \SGol$ for every $j \in \N$, then $\GRS_\gamma^n((a,b),m)$ is a pair of order $\ord a$ and length $2^n m$ for each $n\in\N$.
\item If $(a,b)$ is unimodular (resp., binary), $\len a=\len b=m$, $\ord a=\ord b$, and $\gamma_j \in \UGol$ (resp., $\BGol$) for every $j \in \N$, then $\GRS_\gamma^n((a,b),m)$ is a unimodular (resp., binary) pair of order $\ord a$ and length $2^n m$ for each $n \in \N$.
\end{enumerate}
\end{construction}
\begin{proof}
Each claim is proved inductively using \cref{GRS} along with \cref{Zachary} (for the first claim), \cref{Samantha} (for the second and third claims), and \cref{Ulysses} (resp., \cref{Barbara}) for the parts of the fourth claim not already established in the third claim.
\end{proof}
\begin{example}\label{Hans}
We consider some examples of how \cref{IGRS} can be used to construct Golay pairs.
To easily encode Golay pairs, we represent each binary sequence $f$ of length $2^n$ (for $n \in \N$) as a Boolean function in $n$ variables.
This is a very convenient formalism developed by Davis and Jedwab \cite{Davis-Jedwab} that captures all the known Golay pairs whose lengths are powers of $2$; these pairs were already known to Golay \cite[pp.~85--86]{Golay-61}, who obtained them via a closely related formalism.
We use the same idea as these previous authors, but with a slightly more convenient indexing.
We let $\Ftwo$ be the binary field, and write a Boolean function from $\Ftwo^n$ to $\Ftwo$ as a polynomial $F(x_0,x_1,\ldots,x_{n-1}) \in \Ftwo[x_0,x_1,\ldots,x_{n-1}]$, where $x_0,x_1,\ldots,x_{n-1}$ are $n$ indeterminates, none of which appears with a power greater than $1$ in our polynomial $F(x_0,x_1,\ldots,x_{n-1})$.
We use the convention that if we evaluate $F$ with inputs from $\Z$, we mean that those elements should be reduced modulo $2$ (and so mapped into $\Ftwo$) before evaluation.
With this convention, the sequence $f=\sum_{j \in \Z} f_j z^j$ associated to $F$ has $f_j=(-1)^{F(j_0,j_1,\ldots,j_{n-1})}$ whenever $j_0,\ldots,j_{n-1} \in \{0,1\} \subseteq \Z$ with $j=\sum_{u=0}^{n-1} j_u 2^u$ (and $f_j=0$ when $j \not\in \{0,1,\ldots,2^n-1\}$).
Note this concept even works for the the sequences $1$ and $-1$ of length $1$: these correspond to the Boolean functions in zero variables (i.e., constants) $F=0$ and $G=1$, respectively.
Note that negation (resp., substitution of $-z$ for $z$, conjugate reversal) of the binary sequence of length $2^n$ associated with Boolean function $F(x_0,\ldots,x_{n-1})$ changes it into the binary sequence of length $2^n$ associated with Boolean function $F(x_0,\ldots,x_{n-1})+1$ (resp., $F(x_0,\ldots,x_{n-1})+x_0$, $F(x_0+1,\ldots,x_{n-1}+1)$).

If $(a,b)$ is a binary Golay pair of length $2^n$ whose sequences correspond to Boolean functions $A(x_0,\ldots,x_{n-1})$ and $B(x_0,x_1,\ldots,x_{n-1})$, respectively, then it is not hard to show that $\GRS((a,b),2^n)$ is the binary Golay pair of length $2^{n+1}$ whose sequences correspond to the Boolean functions
\begin{align*}
& (1-x_n) A(x_0,\ldots,x_{n-1}) + x_n B(x_0,\ldots,x_{n-1})\text{ and} \\
& (1-x_n) A(x_0,\ldots,x_{n-1}) + x_n B(x_0,\ldots,x_{n-1})+x_n,
\end{align*}
respectively.
Therefore, if we start with $(a,b)$ equal to the binary Golay pair $(1,1)$ of length $1$, and if we let $\gamma=(\gamma_0,\gamma_1,\ldots)$ where every $\gamma_j$ is the identity transformation, then it is not hard to use induction to show that $\GRS_\gamma^n((a,b),1)$ is the binary Golay pair of length $2^n$ whose sequences correspond to the Boolean functions $0$ and $0$ (if $n=0$) or
\begin{align*}
& \sum_{j=0}^{n-2} x_j x_{j+1} \text{ and} \\
& \sum_{j=0}^{n-2} x_j x_{j+1} + x_{n-1}
\end{align*}
(if $n\geq 1$), where we construe empty sums as $0$ when $n=1$.
These are the Boolean functions associated to the original sequence pairs of Golay and Shapiro.

On the other hand, we can obtain very different sequences starting from the same initial pair $(1,1)$ if we use some non-identity transformations between the stages of Golay concatenation.
Now we let $\gamma_j$ be the identity map for all even $j$, but let $\gamma_j=\crev$ for all odd $j$.
Then it is not hard to show by induction that if $(a,b)$ is the binary Golay pair $(1,1)$ of length $1$, then $\GRS_\gamma^n((a,b),1)$ is the binary Golay pair of length $2^n$ associated to the Boolean functions $0$ and $0$ (for $n=0$), $0$ and $x_0$ (for $n=1$), or $C(x_0,\ldots,x_{n-1})$ and $D(x_0,\ldots,x_{n-1})$ (for $n \geq 2$), where 
\begin{multline*}
C(x_0,\ldots,x_{n-1}) = x_0 x_1 + \sum_{j=0}^{\floor{(n-3)/2}} x_{2 j+1} x_{2 j+2} + \sum_{j=0}^{\floor{(n-4)/2}} x_{2 j} x_{2 j+3} \\ +x_0 + x_{2 \floor{(n-1)/2}},
\end{multline*}
and
\[
D(x_0,\ldots,x_{n-1}) = C(x_0,\ldots,x_{n-1}) + x_{n-2+(-1)^n},
\]
and we construe empty sums as $0$ when $n$ is small.
\end{example}
We are interested in the autocorrelation and crosscorrelation demerit factors of $\GRS_\gamma^n((a,b),m)$ as a function of the inputs.
We begin with a useful technical result.
\begin{lemma}\label{Nestor}
Let $(a,b)$ be a isoenergetic Golay pair with $a,b\not=0$.
Then
\[
\frac{|a|^4_0 + |b|^4_0}{(|a|^2_0+|b|^2_0)^2} = \frac{\ADF(a)+1}{2} = \frac{\ADF(b)+1}{2} = 1-\frac{\CDF(a,b)}{2}.
\]
\end{lemma}
\begin{proof}
Since $(a,b)$ is isoenergetic we have $|a|^2_0=|b|^2_0$, and \cref{Horatio} shows that we have $\ADF(a)=\ADF(b)$, and so
\begin{align*}
\frac{|a|^4_0 + |b|^4_0}{(|a|^2_0+|b|^2_0)^2}
& = \frac{|a|^4_0}{4 (|a|^2_0)^2} + \frac{|b|^4_0}{4 (|b|^2_0)^2} \\
& = \frac{\ADF(a)+1}{4} + \frac{\ADF(b)+1}{4} \\
& = \frac{\ADF(a)+1}{2},
\end{align*}
where we used \eqref{Vera} in the second equality.
Since $(a,b)$ is a Golay pair with $\ADF(a)=\ADF(b)$, \cref{Gabriel} shows that $\ADF(a)=1-\CDF(a,b)$, and substituting this in our expression completes the proof.
\end{proof}
Now we show the effect of one step of \cref{IGRS} on the demerit factors.
\begin{proposition}\label{Barney}
Let $(a,b)$ be a isoenergetic Golay pair with $\ord a=\ord b\not=\infty$, and let $m$ be an integer with $\max(\len a,\len b) \leq m$.
Let $\sigma\in\SGol$.
If $(f,g)=\sigma(\GRS((a,b),m))$, then $(f,g)$ is a isoenergetic Golay pair with $\ord(f,g)=\ord(a,b)\not=\infty$, $\max(\len f,\len g) \leq 2 m$, $\ADF(f)=\ADF(g)$, and 
\begin{align*}
\ADF(f)-\frac{1}{3} & = -\frac{1}{2}\left(\ADF(a)-\frac{1}{3}\right) \\
\CDF(f,g)-\frac{2}{3} & = -\frac{1}{2}\left(\CDF(a,b)-\frac{2}{3}\right).
\end{align*}
\end{proposition}
\begin{proof}
First of all, note that $m>0$ since the condition that $\ord(a,b)\not=\infty$ gives $a$ and $b$ positive lengths.
Let $(c,d)=\GRS((a,b),m)$, so that $(f,g)=\sigma(c,d)$.
From \cref{GRS} and we know that $(c,d)$ is a Golay pair with $\ord(c,d)=\ord(a,b)\not=\infty$, and then we can see that $(c,d)=(a+z^m b,a-z^m b)$ has $\max(\len c,\len d) \leq 2 m$.
Then by \cref{Zachary} and \cref{Samantha}, we see that $(f,g)$ is a Golay pair with $\ord(f,g)=\ord(a,b)\not=\infty$ and $\max(\len f,\len g) \leq 2 m$.

Since $(c,d)=\GRS((a,b),m)=(a, z^m b)\ltimes(1,1)$, we apply \cref{Hugo} and \cref{Roger} and extract the constant coefficients to obtain
\begin{align*}
|c|^2_0 & = |a|^2_0 + |b|^2_0 + 2\Re(z^{-m}a\conj{b})_0 \\
|d|^2_0 & = |a|^2_0 + |b|^2_0 - 2\Re(z^{-m}a\conj{b})_0 \\
|c d|^2_0 & = |a|^4_0 + |b|^4_0 -2\Re((z^{-m}a\conj{b})^2)_0.
\end{align*}
Since $a$ and $b$ are of length $m$ or less and $\ord a=\ord b$, only monomials of negative degree occur in $z^{-m}a\conj{b}$, so there is no constant term in $\Re((z^{-m}a\conj{b})^k)$ for any $k > 0$.
Thus we have
\begin{align*}
|c|^2_0 & = |a|^2_0 + |b|^2_0 \\
|d|^2_0 & = |a|^2_0 + |b|^2_0 \\
|c d|^2_0 & = |a|^4_0 + |b|^4_0.
\end{align*}
In view of \eqref{Dorothy}, the first two equations show that $(c,d)$ is isoenergetic, so \cref{Horatio} shows that $\ADF(c)=\ADF(d)$.
Furthermore we use \eqref{Bart} and the expressions above with \cref{Nestor} to see that
\begin{align*}
\CDF(c,d)
& = \frac{|a|^4_0 + |b|^4_0}{(|a|^2_0+|b|^2_0)^2} \\
& = 1-\frac{\CDF(a,b)}{2},
\end{align*}
which means that
\[
\CDF(c,d)-\frac{2}{3} = -\frac{1}{2}\left(\CDF(a,b)-\frac{2}{3}\right).
\]
Since $(c,d)$ is a Golay pair with $\ADF(c)=\ADF(d)$, \cref{Gabriel} tells us that $\CDF(c,d)+\ADF(c)=1$, and similarly $\CDF(a,b)+\ADF(a)=1$ because $\ADF(a)=\ADF(b)$ by \cref{Horatio}, so we can negate both sides of the last equation to obtain
\[
\ADF(c)-\frac{1}{3} = -\frac{1}{2}\left(\ADF(a)-\frac{1}{3}\right).
\]
Since $(c,d)$ is isoenergetic and $\ADF(c)=\ADF(d)$, \cref{Cameron} shows that $(f,g)=\sigma(c,d)$ is isoenergetic, that $\ADF(f)=\ADF(g)=\ADF(c)$, and that $\CDF(f,g)=\CDF(c,d)$.
\end{proof}
Now that we have seen how a single step of \cref{IGRS} transforms the $\ADF$ and $\CDF$ of its input pair, we shall now investigate the effect of multiple steps.
\begin{theorem}\label{Sally}
Let $(f,g)$ be a isoenergetic Golay pair with $\ord f=\ord g\not=\infty$, let $m$ be an integer with $\max(\len f,\len g)\leq m$, and let $\gamma=(\gamma_0,\gamma_1,\ldots)$ be a sequence of transformations from $\SGol$.
Let $(f^{(n)},g^{(n)})=\GRS_\gamma^n((f,g),m)$ for each $n \in \N$.
Then for each $n \in \N$, the pair $(f^{(n)},g^{(n)})$ is a isoenergetic Golay pair with $\ord(f^{(n)},g^{(n)})=\ord(f,g)\not=\infty$, $\max(\len f^{(n)},\len g^{(n)}) \leq 2^n m$, $\ADF(f^{(n)})=\ADF(g^{(n)})$, and
\begin{align*}
\ADF(f^{(n)})-\frac{1}{3} & = \left(-\frac{1}{2}\right)^n \left(\ADF(f)-\frac{1}{3}\right) \\
\CDF(f^{(n)},g^{(n)})-\frac{2}{3} & = \left(-\frac{1}{2}\right)^n\left(\CDF(f,g)-\frac{2}{3}\right)
\end{align*}
for each $n \in \N$.
Thus
\begin{align*}
\lim_{n\to\infty} \ADF(f^{(n)})=\lim_{n\to\infty} \ADF(g^{(n)}) & = \frac{1}{3}, \\
\lim_{n\to\infty} \CDF(f^{(n)},g^{(n)}) & = \frac{2}{3}.
\end{align*}
\end{theorem}
\begin{proof}
The asymptotic results follow immediately from the preceding formulae for the demerit factors, and the proof of all the non-asymptotic results proceeds by induction on $n$.
If $n=0$, then \cref{Horatio} shows that $\ADF(f)=\ADF(g)$, and then Lemmas \ref{Zachary} and \ref{Cameron} and \cref{Samantha} show that $(f^{(0)},g^{(0)})=\gamma_0(f,g)$ is a isoenergetic Golay pair with $\ord(f^{(0)},g^{(0)})=\ord(f,g)\not=\infty$, $\max(\len f^{(0)},\len g^{(0)})=\max\{\len f,\len g\}\leq m$, $\ADF(f^{(0)})=\ADF(g^{(0)})=\ADF(f)=\ADF(g)$, and $\CDF(f^{(0)},g^{(0)})=\CDF(f,g)$.
If $n>0$ and we assume that our conclusions hold for $(f^{(n-1)},g^{(n-1)})$, then \cref{Barney} shows that our conclusions hold for $(f^{(n)},g^{(n)})$.
\end{proof}
\begin{remark}\label{Genevieve}
Let $m$ be a positive integer and let $(f,g)$ be be a unimodular Golay pair of length $m$ with $\ord f=\ord g$.
Let $\gamma=(\gamma_0,\gamma_1,\ldots)$ be a sequence of transformations from $\UGol$.
Then $(f,g)$, $m$, and $\gamma$ satisfy the hypotheses of \cref{Sally}, and in addition to the conclusions therefrom, we also know that for each $n\in\N$ the pair $(f^{(n)},g^{(n)})$ is unimodular and length $2^n m$.
To see this, note that $(f,g)$ is isoenergetic by \cref{Felicia}, that $f$ and $g$ are of positive length and equal order, so $\ord f=\ord g\not=\infty$, and that $\UGol$ is a subgroup of $\SGol$, so that $(f,g)$, $m$, and $\gamma$ satisfy all the hypotheses of \cref{Sally}, while \cref{IGRS} shows that $(f^{(n)},g^{(n)})$ is unimodular and of length $2^n m$ for every $n\in \N$.
\end{remark}
\begin{remark}
Let $m$ be a positive integer and let $(f,g)$ be be a binary Golay pair of length $m$ with $\ord f=\ord g$.
Let $\gamma=(\gamma_0,\gamma_1,\ldots)$ be a sequence of transformations from $\BGol$.
Then $(f,g)$, $m$, and $\gamma$ satisfy the hypotheses of \cref{Sally}, and in addition to the conclusions therefrom, we also know that for each $n\in\N$ the pair $(f^{(n)},g^{(n)})$ is binary and length $2^n m$.
To see this, note that binary sequences are unimodular and $\BGol$ is a subgroup of $\UGol$, so we may use \cref{Genevieve}, and then \cref{IGRS} shows that the pair $(f^{(n)},g^{(n)})$ is binary for every $n\in \N$.
\end{remark}

\section{An Iterated Golay Interleaving Construction}\label{Valerie}

In this section, we shall explore how an iterated interleaving construction for Golay pairs influences demerit factors.
In the previous section, we looked at the Golay-Rudin-Shapiro construction, which was noted to be a special case of Golay's concatenation construction.
We present the analogous special case of Golay's interleaving (Construction \ref{GIC}), where we use $1$ for the parameter $m$ and $(1,1)$ for the first pair $(a,b)$ in that construction.
\begin{construction}[Simple Golay interleaving]\label{SGI}
If $(a,b) \in \laur^2$ with $\ord a+\deg a=\ord b+\deg b$, then we set 
\begin{align*}
\SGI(a,b)
& =\scal_{1,z^{2 (\ord a+\deg a)}}\left((\scal_{1,z}(1,1))\ltimes(\subs_{z^2}(a,b))\right)\\
& =\left(a(z^2)+z b(z^2),b^\ddag(z^2)-z a^\ddag(z^2)\right).
\end{align*}
We prove some properties of our construction.
\begin{enumerate}[(i).]
\item If $(a,b)$ is a Golay pair, then so is $\SGI(a,b)$.
\item If $\len a=\len b$, then $\SGI(a,b)$ is a pair of order $2\ord a$ and length $2\len a$.
\item If $(a,b)$ is a unimodular (resp., binary) pair with $\len a=\len b$, then $\SGI(a,b)$ is a unimodular (resp., binary) pair of length $2 \len a$.
\end{enumerate}
\end{construction}
\begin{proof}
The claim that $\SGI(a,b)$ is a Golay pair whenever $(a,b)$ is follows from \cref{Zachary} and \cref{Ursula} and the fact that $(1,1)$ is a Golay pair.
In the second and third claims, the additional assumption that $\len a=\len b$ along with the original assumption $\ord a+\deg a=\ord b+\deg b$ imply that $\ord a=\ord b$, and then the second expression for $\SGI(a,b)$ makes clear that $\ord\SGI(a,b)=2\ord a$ and $\len\SGI(a,b)=2\len a$ and that $\SGI(a,b)$ is unimodular (resp., binary) if $(a,b$ is unimodular (resp., binary).
\end{proof}
We now iterate the previous construction, also allowing for the application of transformations from $\SGol$.
\begin{construction}[Iterated simple Golay interleaving]\label{ISGI}
Let $(a,b)$ be a pair in $\laur^2$ with $\len a=\len b$ and $\ord a=\ord b$ and let $\gamma=(\gamma_0,\gamma_1,\ldots)$ be a sequence of transformations from $\SGol$.
For $n \in \N$, we define $\SGI_\gamma^n(a,b)$ recursively by setting $\SGI_\gamma^0(a,b) =\gamma_0(a,b)$ and for $n > 0$ setting $\SGI_\gamma^{n+1}=\gamma_{n+1}(\SGI(\SGI_\gamma^n(a,b)))$.
We prove some properties of our iterated construction.
\begin{enumerate}[(i).]
\item If $(a,b)$ is a Golay pair, then so is $\SGI_\gamma^n(a,b)$ for every $n \in \N$.
\item $\SGI_\gamma^n(a,b)$ is a pair of length $2^n \len a$ and order $2^n \ord a$ for every $n\in \N$.
\item If $(a,b)$ is a unimodular (resp., binary) pair and $\gamma_j \in \UGol$ (resp., $\gamma_j \in \BGol$) for every $j \in \N$, then $\SGI_\gamma^n(a,b)$ is a unimodular (resp., binary) pair of length $2^n \len a$ and order $2^n \ord a$ for every $n \in \N$.
\end{enumerate}
\end{construction}
\begin{proof}
Throughout this proof, it is useful to note that if two sequences, $f$ and $g$, meet any two of the three conditions (I) $\len f=\len g$, (II) $\ord f=\ord g$, or (III) $\ord f+\deg f=\ord g+\deg g$, then they must also meet the third condition.
This is clear because the pair $(f,g)=(0,0)$ meets all three conditions, a pair consisting of one zero sequence and one nonzero sequence meets neither (I) nor (II), and a pair $(f,g)$ of two nonzero sequences has $\len f=\deg f-\ord f+1$ and $\len g=\deg g-\ord g+1$.
  
For pairs of sequences with matching lengths and orders, transformations from $\SGol$ preserve the length and order (see \cref{Samantha}) while $\SGI$ doubles the length and order (see \cref{SGI}), so one can inductively establish the second claim for each $n \in \N$, which allows the recursive construction to carry on to the next step, since the input of $\SGI$ must be a pair $(f,g)$ with $\ord f+\deg f=\ord g+\deg g$.
The third claim follows from the second along with an induction using the final result in \cref{SGI} along with \cref{Ulysses} (for unimodular sequences) or \cref{Barbara} (for binary sequences).
The first claim is proved by induction using \cref{Zachary} and \cref{SGI}.
\end{proof}
\begin{example}
We consider how \cref{ISGI} can be used to construct Golay pairs, using the same Boolean function formalism described in \cref{Hans}.
If $(a,b)$ is a binary Golay pair of length $2^n$ whose sequences correspond to Boolean functions $A(x_0,\ldots,x_{n-1})$ and $B(x_0,x_1,\ldots,x_{n-1})$, respectively, then it is not hard to show that $\SGI(a,b)$ is the binary Golay pair of length $2^{n+1}$ whose sequences correspond to the Boolean functions
\begin{align*}
& (1-x_0) A(x_1,\ldots,x_n) + x_0 B(x_1,\ldots,x_n)\text{ and} \\
& (1-x_0) B(x_1+1,\ldots,x_n+1) + x_0 A(x_1+1,\ldots,x_n+1)+x_0,
\end{align*}
respectively.
Therefore, if we start with $(a,b)$ equal to the binary Golay pair $(1,1)$ of length $1$, and if we let $\gamma=(\gamma_0,\gamma_1,\ldots)$ where every $\gamma_j$ is the identity transformation, then it is not hard to use induction to show that $\SGI_\gamma^n(a,b)$ is the binary Golay pair of length $2^n$ whose sequences correspond to the Boolean functions $0$ and $0$ (if $n=0$), $0$ and $x_0$ (if $n=1$), or
\begin{align*}
& x_{n-2} x_{n-1} + \sum_{j=0}^{n-3} x_j x_{j+2} + \sum_{j=0}^{n-3} x_j \text{ and} \\
& x_{n-2} x_{n-1} + \sum_{j=0}^{n-3} x_j x_{j+2} + \sum_{j=0}^{n-3} x_j + x_1 + 1
\end{align*}
(for $n\geq 2$), where we construe empty sums in these expressions as $0$ when $n=2$.

On the other hand, we can obtain very different sequences starting from the same initial pair $(1,1)$ if we use some non-identity transformations between the stages of simple Golay interleaving.
Recall from \cref{Hans} that conjugate reversal of the binary sequence of length $2^n$ associated with Boolean function $F(x_0,\ldots,x_{n-1})$ changes it into the binary sequence of length $2^n$ associated with Boolean function $F(x_0+1,\ldots,x_{n-1}+1)$.
Now we let $\gamma_j$ be the identity map for all even $j$, but let $\gamma_j=\crev$ for all odd $j$.
Then with some care one can prove by induction that if $(a,b)$ is the binary Golay pair $(1,1)$ of length $1$, then $\SGI_\gamma^n(a,b)$ is the binary Golay pair of length $2^n$ associated to the Boolean functions $0$ and $0$ (for $n=0$), $0$ and $x_0$ (for $n=1$), $x_0 x_1$ and $x_0 x_1+x_1+1$ (for $n=2$), or $C(x_0,\ldots,x_n)$ and $D(x_0,\ldots,x_n)$ (for $n\geq 3$), where
\begin{multline*}
C(x_0,\ldots,x_n) = x_{n-3} x_{n-1} + \sum_{j=0}^{\floor{(n-2)/2}}  x_{n-2-2 j} x_{n-1-2 j} \\ + \sum_{j=0}^{\floor{(n-5)/2}}  x_{n-5-2 j} x_{n-2-2 j} + \sum_{j=0}^{n-4} x_j + x_{n-2} + (n+1) x_0 + \binom{n-1}{2},
\end{multline*}
and
\[
D(x_0,\ldots,x_n) = C(x_0,\ldots,x_n)+n x_0 + (n+1) x_2,
\]
and we construe empty sums as $0$ when $n$ is small.
\end{example}
Now we investigate how a single step of \cref{ISGI} affects demerit factors.
\begin{proposition}\label{Trish}
Let $(a,b)$ be a isoenergetic Golay pair with $\len a=\len b>0$ and $\ord a=\ord b$.
Let $\sigma \in \SGol$.
If $(f,g) = \sigma(\SGI(a,b))$, then $(f,g)$ is a isoenergetic Golay pair with $\len(f,g)=2\len(a,b)>0$, $\ord(f,g)=2\ord(a,b)$, $\ADF(f)=\ADF(g)$, and
\begin{align*}
\ADF(f)-\frac{1}{3} & = -\frac{1}{2}\left(\ADF(a)-\frac{1}{3}\right) + W(a,b)\\
\CDF(f,g)-\frac{2}{3} & = -\frac{1}{2}\left(\CDF(a,b)-\frac{2}{3}\right) - W(a,b),
\end{align*}
where
\[
W(a,b) = \frac{2\Re(\conj{z}(a\conj{b})^2)_0}{(|a|^2_0 + |b|^2_0)^2}.
\]
\end{proposition}
\begin{proof}
Let $(c,d) = \SGI(a,b)$, so that $(f,g)=\sigma(c,d)$.
From \cref{SGI} we know that $(c,d)$ is a Golay pair with $\len(c,d)=2\len(a,b)>0$ and $\ord(c,d)=2\ord(a,b)$.
Then by \cref{Zachary} and \cref{Samantha}, we can see that $(f,g)=\sigma(c,d)$ is a Golay pair with $\len(f,g)=2\len(a,b)>0$ and $\ord(f,g)=2\ord(a,b)$.

Let $(c',d')=(1,z)\ltimes(a(z^2),b(z^2))$ so that $(c,d)=\scal_{1,z^{2 (\ord a+\deg a)}}(c',d')$.
Because $|z|^2=1$, we see that $|c|^2$, $|d|^2$, and $|c d|^2$ are the same as $|c'|^2$, $|d'|^2$, and $|c' d'|^2$, respectively, and we can calculate these by using Lemmas \ref{Hugo} and \ref{Roger} on $(c',d')$ to obtain
\begin{align*}
|c|^2_0 & = |a(z^2)|^2_0 + |b(z^2)|^2_0 + 2\Re\left(\conj{z}a(z^2)\conj{b(z^2)}\right)_0\\
|d|^2_0 & = |a(z^2)|^2_0 + |b(z^2)|^2_0 - 2\Re\left(\conj{z}a(z^2)\conj{b(z^2)}\right)_0\\
|cd|^2_0 & = |a(z^2)|^4_0 + |b(z^2)|^4_0 -2\Re\left(\left(\conj{z}a(z^2)\conj{b(z^2)}\right)^2\right)_0.
\end{align*}
Notice that only monomials of odd degree occur in $\conj{z}a(z^2)\conj{b(z^2)}$, so there is no constant term in it or its conjugate.
Thus the final term in each of the first two expressions vanishes.
Further, we note that $\Re((\conj{z}a(z^2)\conj{b(z^2)})^2)_0 = \Re(\conj{z}(a\conj{b})^2)_0$.
As such,
\begin{align*}
|c|^2_0 & = |a|^2_0 + |b|^2_0\\
|d|^2_0 & = |a|^2_0 + |b|^2_0\\
|cd|^2_0 & = |a|^4_0 + |b|^4_0 -2\Re\left(\conj{z}(a\conj{b})^2\right)_0.
\end{align*}
In view of \eqref{Dorothy}, the first two equations show that $(c,d)$ is isoenergetic, so \cref{Horatio} shows that $\ADF(c)=\ADF(d)$.
Now we use \eqref{Bart} and the expressions above to see that
\[
\CDF(c,d) = \frac{|a|^4_0 + |b|^4_0}{\left(|a|^2_0 + |b|^2_0\right)^2} - \frac{2\Re\left(\conj{z}(a\conj{b})^2\right)_0}{\left(|a|^2_0 + |b|^2_0\right)^2}.
\]
Then use \cref{Nestor} and the definition of $W(a,b)$ in the statement of this lemma to see that
\[
\CDF(c,d) = 1-\frac{\CDF(a,b)}{2} - W(a,b),
\]
and then we can subtract $2/3$ from both sides to obtain
\[
\CDF(c,d)-\frac{2}{3} = -\frac{1}{2}\left(\CDF(a,b)-\frac{2}{3}\right) - W(a,b).
\]
Since $(c,d)$ is a Golay pair with $\ADF(c)=\ADF(d)$, \cref{Gabriel} tells us that $\CDF(c,d)+\ADF(c)=1$, and similarly $\CDF(a,b)+\ADF(a)=1$ because $\ADF(a)=\ADF(b)$ by \cref{Horatio}, so we can negate both sides of the last equation to obtain
\[
\ADF(c)-\frac{1}{3} = -\frac{1}{2}\left(\ADF(a)-\frac{1}{3}\right) + W(a,b).
\]
Since $(c,d)$ is isoenergetic and $\ADF(c)=\ADF(d)$, \cref{Cameron} shows that $(f,g)=\sigma(c,d)$ is isoenergetic, that $\ADF(f)=\ADF(g)=\ADF(c)$, and that $\CDF(f,g)=\CDF(c,d)$.
\end{proof}
We now wish to investigate the autocorrelation and crosscorrelation demerit factors of $\SGI_\gamma^n(a,b)$ from \cref{ISGI} by applying the last proposition multiple times, but the term $W(a,b)$ that appears in that proposition can make calculations troublesome.
We find that if we restrict $\sigma$ to come from a carefully selected subgroup of $\SGol$, then when we use the output pair $(f,g)$ from the proposition as an input into the same proposition (for the next step of \cref{ISGI}), we will find that $W(f,g)=0$, thus simplifying the calculation.
We first describe the subgroup of $\SGol$ that allows for this simplification.
\begin{definition}[Restricted Golay Group $\RGol$]\label{Richard}
The {\it restricted Golay group}, written $\RGol$, is the subgroup of $\SGol$ generated by $\swap$, $\crev\circ\crev'$, $\srev$, all transformations $\scal_{u,v}$ where $u,v$ are nonzero complex numbers with $|u|=|v|$, and all transformations $\subs_{w z}$, where $w$ is a unimodular complex number.
\end{definition}
\begin{remark}
The only difference between the generating set of $\RGol$ and that of $\SGol$ is that $\RGol$ has the single generator $\crev\circ\crev'$ in place of the two generators $\crev$ and $\crev'$ for $\SGol$.
Thus $\RGol$ is a subgroup of $\SGol$, and indeed a proper subgroup because it is straightforward to show that if $(a,b)=(1+z+z^2-z^3,1+z-z^2+z^3)$ and $(f,g)=\gamma(a,b)$ for some $\gamma\in\RGol$, then $f g$ has terms of both even and odd degree, but $a^\ddag b$ has only terms of even degree, so $\gamma(a,b)=(f,g)\not=(a^\dag,b)=\crev(a,b)$, and so $\crev\in\SGol\smallsetminus\RGol$.
\end{remark}
\begin{definition}[Restricted Unimodular Golay Group $\RUGol$]
The {\it restricted unimodular Golay group}, written $\RUGol$, is $\RGol\cap\UGol$.
\end{definition}
\begin{remark}
Note that $\RUGol$ contains $\swap$, $\crev\circ\crev'$, $\srev$, all transformations $\scal_{u,v}$ where $u,v$ are unimodular complex numbers, and all transformations $\subs_{w z}$, where $w$ is a unimodular complex number, because all these transformations are in both $\RGol$ and $\UGol$.
\end{remark}
\begin{definition}[Restricted Binary Golay Group $\RBGol$]
The {\it restricted binary Golay group}, written $\RBGol$, is $\RGol\cap\BGol$.
\end{definition}
\begin{remark}
Note that $\RBGol$ contains $\swap$, $\crev\circ\crev'$, $\srev$, $\scal_{-1,1}$, $\scal_{1,-1}$, and $\subs_{-z}$, because all these transformations are in both $\RGol$ and $\BGol$.
Recall that $\crev\circ\crev'$ and $\srev$ have the same effect when applied to sequences with real terms.
\end{remark}
The following lemma shows that if we use a transformation $\sigma$ from $\RGol$ in \cref{Trish}, then the output pair $(f,g)$ from that proposition has the property that $W(f,g)=0$ when we use $(f,g)$ again as an input pair for that proposition.
\begin{lemma}\label{Jesse}
Let $(a,b) \in \laur^2$ with $\ord a+\deg a=\ord b+\deg b$, let $\sigma\in\RGol$, and let $(f,g)=\sigma(\SGI(a,b))$.  Then $\Re(\conj{z} (f\conj{g})^2)_0=0$.
\end{lemma}
\begin{proof}
Let $(c,d)=\SGI(a,b)$, so that $(f,g)=\sigma(c,d)$.
From the formula for $\SGI(a,b)$ in \cref{SGI}, and using the hypothesis that $\ord a+\deg a=\ord b+\deg b$, we see that
\begin{align*}
c\conj{d}
& =\left(a(z^2)+z b(z^2)\right) \left(z^{-2(\ord b+\deg b)} b(z^2)-z^{-2(\ord a+\deg a)-1} a(z^2)\right) \\
& =-z^{-2(\ord a+\deg a)-1} (a(z^2)^2-z^2 b(z^2)^2),
\end{align*}
which has only odd degree terms.
We say that a Laurent polynomial $x(z)$ is {\it parity-pure} to mean that $x(z)$ does not have both a term of even degree and a term of odd degree, and we say that a pair $(x,y)$ of Laurent polynomials is {\it parity-pure} to mean that the Laurent polynomial $x\conj{y}$ is parity-pure.
So $(c,d)$ is parity-pure.
We claim that $(f,g)=\sigma(c,d)$ is also parity-pure.
Recall the generating set of $\RGol$ specified in \cref{Richard}.
Since this set of generators is closed under inversion (see \cref{Gertrude}, and note that $\crev\circ\crev'$ is an involution), $\sigma$ is a composition of some of these generators.
We claim that any such generator $\eta$, when applied to a parity-pure pair $(x,y) \in\laur^2$ , produces another parity-pure pair $(x',y')=\eta(x,y)$, because of the following considerations.
\begin{itemize}
\item If $\eta=\swap$, then $x'\conj{y'}=y\conj{x}=\conj{x\conj{y}}$, which is parity-pure.
\item If $\eta=\crev\circ\crev'$, then $x'\conj{y'}=x^\ddag\conj{y^\ddag}=z^{\ord x+\deg x-\ord y-\deg y} \conj{x \conj{y}}$, which is parity-pure.
\item If $\eta=\srev$, then
\begin{align*}
x'\conj{y'}
&=z^{\ord x+\deg x-\ord y-\deg y} x(z^{-1})\conj{y(z^{-1})}\\
&=z^{\ord x+\deg x-\ord y-\deg y} (x\conj{y})(z^{-1}),
\end{align*}
which is parity-pure.
\item If $\eta=\scal_{u,v}$ where $u$ and $v$ are nonzero complex numbers, then $x'\conj{y'}=u\conj{v} x\conj{y}$, which is parity-pure.
\item If $\eta=\subs_{w z}$ where $w$ is a unimodular complex number, then $x'\conj{y'}=x(w z)\conj{y(w z)}$.  But if we let $h=\conj{y}$, then $\conj{y(w z)}=h(\conj{w}^{-1} z)=h(w z)$, so that $x'\conj{y'}=(x\conj{y})(w z)$, which is parity-pure.
\end{itemize}
Thus we conclude that $(f,g)=\sigma(c,d)$ is parity-pure, and so $\conj{z} (f \conj{g})^2$ has only terms of odd degree, so that $\Re(\conj{z} (f\conj{g})^2)_0=0$.
\end{proof}
Now we can use \cref{Trish} with \cref{Jesse} to analyze demerit factors of pairs produced by \cref{ISGI}.
\begin{theorem}\label{Griffin}
Let $(f,g)$ be a isoenergetic Golay pair with $\len f=\len g>0$ and $\ord f=\ord g$.
Let $\gamma=(\gamma_0,\gamma_1,\ldots)$ be a sequence of transformations from $\RGol$.
Let $(f^{(n)},g^{(n)})=\SGI^n_\gamma(f,g)$ for each $n \in \N$.
Then for each $n\in\N$, the pair $(f^{(n)},g^{(n)})$ is a isoenergetic Golay pair with $\len(f^{(n)},g^{(n)})=2^n\len(f,g)>0$, $\ord(f^{(n)},g^{(n)})=2^n \ord(f,g)$, and $\ADF(f^{(n)})=\ADF(g^{(n)})$.
We have $\ADF(f^{(0)})=\ADF(f)$ and $\CDF(f^{(0)},g^{(0)})=\CDF(f,g)$, and for $n > 0$ we have
\begin{align*}
\ADF(f^{(n)})-\frac{1}{3} & = \left(-\frac{1}{2}\right)^n \left(\ADF(f)-\frac{1}{3}\right) + \left(-\frac{1}{2}\right)^{n-1}W_0\\
\CDF(f^{(n)},g^{(n)})-\frac{2}{3} & = \left(-\frac{1}{2}\right)^n\left(\CDF(f,g)-\frac{2}{3}\right) - \left(-\frac{1}{2}\right)^{n-1}W_0,
\end{align*}
where
\[
W_0 = \frac{2\Re((\conj{z}(f^{(0)}\conj{g^{(0)}})^2)_0}{(|f^{(0)}|^2_0 + |g^{(0)}|^2_0)^2}.
\]
Thus
\begin{align*}
\lim_{n\to\infty} \ADF(f^{(n)})=\lim_{n\to\infty} \ADF(g^{(n)}) & = \frac{1}{3}, \\
\lim_{n\to\infty} \CDF(f^{(n)},g^{(n)}) & = \frac{2}{3}.
\end{align*}
\end{theorem}
\begin{proof}
The asymptotic results follow immediately from the formulae for the demerit factors, and the proof of all the non-asymptotic results proceed by induction on $n$.
If $n=0$, then \cref{Horatio} shows that $\ADF(f)=\ADF(g)$, and then \cref{Zachary}, \cref{Cameron}, and \cref{Samantha} show that $(f^{(0)},g^{(0)})$ is a isoenergetic Golay pair with $\len(f^{(0)},g^{(0)})=\len(f,g)>0$, $\ord(f^{(0)},g^{(0)})=\ord(f,g)$, $\ADF(f^{(0)})=\ADF(g^{(0)})=\ADF(f)=\ADF(g)$, and $\CDF(f^{(0)},g^{(0)})=\CDF(f,g)$.
For $n=1$, \cref{Trish} gives us all the desired results.
If $n > 1$ and we assume the results hold for $(f^{(n-1)},g^{(n-1)})$, then we apply \cref{Trish}, which immediately tells us that $(f^{(n)},g^{(n)})$ is a isoenergetic Golay pair of length $2^n\len(f,g)$ and order $2^n\ord(f,g)$ with $\ADF(f^{(n)})=\ADF(g^{(n)})$.  Furthermore, since \cref{Jesse} tells us that $W(f^{(n-1)},g^{(n-1)})=0$, the demerit factor formulae that \cref{Trish} supplies become
\begin{align*}
\ADF(f^{(n)})-\frac{1}{3} & = -\frac{1}{2}\left(\ADF(f^{(n-1)})-\frac{1}{3}\right) \\
\CDF(f^{(n)},g^{(n)})-\frac{2}{3} & = -\frac{1}{2}\left(\CDF(f^{(n-1)},g^{(n-1)})-\frac{2}{3}\right),
\end{align*}
into which we substitute the values of $\ADF(f^{(n-1)})$ and $\CDF(f^{(n-1)},g^{(n-1)})$ from the induction hypothesis to obtain the desired result.
\end{proof}
\begin{remark}\label{Sophie}
Let $(f,g)$ be a unimodular Golay pair of nonzero sequences with $\ord f=\ord g$.
Let $\gamma=(\gamma_0,\gamma_1,\ldots)$ be a sequence of transformations from $\RUGol$.
Then $(f,g)$ and $\gamma$ satisfy the hypotheses of \cref{Griffin}, and in addition to the conclusions therefrom, we also know that for each $n\in\N$ the pair $(f^{(n)},g^{(n)})$ is unimodular.
To see this, note that \cref{Felicia} shows that $(f,g)$ is isoenergetic and $\len f=\len g$ (which is nonzero since the sequences are nonzero), and note that $\RUGol$ is a subgroup of $\RGol$, so that $(f,g)$ and $\gamma$ satisfy all the hypotheses of \cref{Griffin}, while \cref{ISGI} shows that $(f^{(n)},g^{(n)})$ is unimodular for every $n\in \N$.
\end{remark}
\begin{remark}
Let $(f,g)$ be a binary Golay pair of nonzero sequences with $\ord f=\ord g$.
Let $\gamma=(\gamma_0,\gamma_1,\ldots)$ be a sequence of transformations from $\RBGol$.
Then $(f,g)$ and $\gamma$ satisfy the hypotheses of \cref{Griffin}, and in addition to the conclusions therefrom, we also know that for each $n\in\N$ the pair $(f^{(n)},g^{(n)})$ is binary.
To see this, note that binary sequences are unimodular and $\RBGol$ is a subgroup of $\RUGol$, so we may use \cref{Sophie}, and then \cref{ISGI} shows that the pair $(f^{(n)},g^{(n)})$ is binary for every $n\in \N$.
\end{remark}

\section{Open Problems}\label{Zeke}
In Sections \ref{Taylor} and \ref{Valerie}, we observed that we can find infinitely many families of unimodular (or binary) Golay pairs whose sequences have autocorrelation demerit factors and crosscorrelation demerit factors that tend to $1/3$ and $2/3$, respectively, as the length of the sequences tends to infinity.
So far, we have been unable to find a construction that creates a family of Golay pairs that deviates from these asymptotic demerit factor values.
This leaves us with two open problems about the existence of such a families of Golay pairs.
Recall from \cref{Felicia} that if $p=(f,g)$ is a unimodular Golay pair consisting of nonzero sequences, then $\len f=\len g$ and $\ADF(f)=\ADF(g)$, so we can write $\len p$ and $\ADF(p)$ for the common values.
\begin{problem}
Let $\{p_n\}_{n=1}^\infty$ be a sequence of binary Golay pairs with nonzero sequences such that $\len p_n \to \infty$ as $n \to \infty$.
Must it be true that $\lim_{n\to\infty} \ADF(p_n)=1/3$ and $\lim_{n \to \infty} \CDF(p_n)=2/3$?
\end{problem}
Similarly, there is the open problem for the more general unimodular case.
\begin{problem}
Let $\{p_n\}_{n=1}^\infty$ be a sequence of unimodular Golay pairs with nonzero sequences such that $\len p_n \to \infty$ as $n \to \infty$.
Must it be true that $\lim_{n\to\infty} \ADF(p_n)=1/3$ and $\lim_{n \to \infty} \CDF(p_n)=2/3$?
\end{problem}

\providecommand\noopsort[1]{}

\end{document}